\documentclass[12pt,astron]{thesis}

\usepackage{lettrine}
\usepackage{amsmath}

\usepackage{graphicx}
\usepackage{txfonts}

\usepackage{lscape}
\usepackage{marginnote}
\newcommand{\clearemptydoublepage}{\newpage{\pagestyle{empty}\cleardoublepage}}

\renewcommand{\AA}{\small $\overset{_{\circ}}{\text{A}}$\normalsize}

\def\lesssim{\mathrel{\hbox{\rlap{\hbox{\lower4pt\hbox{$\sim$}}}\hbox{$<$}}}}
\def\gtrsim{\mathrel{\hbox{\rlap{\hbox{\lower4pt\hbox{$\sim$}}}\hbox{$>$}}}}

\newcommand{\tableline}{\noalign{\smallskip}\hline\noalign{\smallskip}}

\usepackage{natbib,twoopt}
\usepackage[breaklinks=true]{hyperref} 
\bibpunct{(}{)}{;}{a}{}{,}             

\usepackage{longtable}
\let\tableline=\hline

\usepackage{pdfpages}

\newcommand\invisiblesection[1]{%
  \refstepcounter{section}%
  \addcontentsline{toc}{section}{\protect\numberline{\thesection}#1}%
  \sectionmark{#1}}

\begin{document}

\frontmatter
\title{A  Journey  across the Hertzsprung-Russell diagramÊ with 3D hydrodynamical simulations of cool stars}

\author{Andrea Chiavassa}
\titulohdr

%
%
%
%

%

\clearemptydoublepage
\thispagestyle{empty}


\clearemptydoublepage
\tableofcontents
\listoffigures

%

\clearemptydoublepage
\mainmatter

\clearemptydoublepage

\chapter*{Preface \label{ch:preface}}

\lettrine[lines=3,nindent=4pt]
{I}{n} this manuscript (\textit{m\'emoire}) I summarise the work I have carried out in the last $\sim$10 years. My research projects are articulated around various fields in astrophysics having, as unifying theme, the 3D modelling and the observation of stellar atmospheres. In particular, my field of research covers all late type stars, from the Main Sequence stars to the advanced stages (red giant and supergiant stars).

My work is based on the quantitative analysis of images, synthetic spectra, bolometric luminosity, planetary transits of different types of stars and on the modeling of their atmospheres. They include the use and development of 3D simunlations that are necessary for the interpretation of the data obtained with the new instruments equipping the space borne and ground telescopes. Interferometry and spectroscopy hold a privileged place in my research.

My research themes concern interdisciplinary aspects by addressing questions related to atmospheric physics in relation to 3D hydrodynamic models. The predictions thus obtained are used to drive a vast campaign of interferometric, astrometric, photometric, spectroscopic and imaging observations. Finally, 3D simulations are also directly used to drive the development of new generation instrument at the VLTI, as well as very large future observatories and are directly linked to the ESA Gaia and PLATO missions.
%

\begin{flushright}
Nice, 7 November 2018. Andrea Chiavassa
\end{flushright}
\clearemptydoublepage

\chapter{Introduction \label{ch:intro}}

\section{Context}
\lettrine[lines=3,nindent=4pt]
{T}{he information} we have to study the stars comes from the photons they emitted. The radiation that escapes from the atmosphere is the only source of information for the study of these objects. In this context, the description of stellar atmospheres is of paramount importance, because it is there that radiation is formed. Convection plays a central role in the structure, dynamics and appearance of evolved stars.
In recent years, it has become possible to produce multidimensional (and in particular three-dimensional, 3D) hydrodynamic simulations of gas motion in the atmospheric layers of stars, coupled with radiation. A 3D approach is required for qualitative and quantitative surface analysis of most stars. The exchange of energy between the gas and the radiation field is essential because it determines the thermal gradient and drains convective movements. 3D models are ab initio, time-dependent, multidimensional and non-local. However, they require considerable computation time and do not yet allow the inclusion of a complex radiative transfer treatment as in the case of 1D.
Understanding the stellar atmosphere is crucial for several reasons:
\begin{itemize}
\item The atmosphere is a window into the stars, and serves as a link between atmospheric models, stellar evolution models, and observations. Stellar evolution phenomena manifest themselves in stellar surfaces as a change in chemical composition and fundamental parameters (e.g. radius, surface gravity, effective temperature and brightness) due to dredge-up processes, internal waves, circulations, rotation. Atmospheric models are therefore necessary as a boundary condition for interior star models and are essential for comparison with observed stellar oscillations (Helio- and Astero-seismology).
\item The atmosphere is also the interface with the interstellar medium. In this respect, the effects on the interstellar medium, through radiation and/or mass loss, reflect the physics of the atmosphere and affect the dynamic and chemical evolution of the Galaxy.
\item Finally, modelling of the stellar atmosphere is also motivated by modernes observations of stellar surfaces, which show a very dynamic landscape and complex structures, especially in the presence of magnetic fields.
\end{itemize}

A realistic modelling taking into account all the richness of the processes at work in the atmosphere (convection, shocks, radiative transfer, ionization, formation of molecules/dust) is essential for the correct analysis of stellar observables. My research develops in this context: I study the atmosphere of late stars (from the Main Sequence to the red supergiant stage) with three-dimensional radiative hydrodynamic models. The key point of my research is the synergy between theory and observations. The 3D numerical simulations are compared to current and future observations (interferometric, photometric, spectroscopic, astrometric and images) using the {{\sc Optim3D}} radiative transfer code, which I developed, to calculate the observables.

\newpage

\section{Open questions}

\lettrine[lines=3,nindent=4pt]
{A}{s} I will do in next chapter, I separate the discussions in two main topics: namely (1) stars from main sequence to Red Gian Branch (RGB) phase and (2) evolved stars such as Red SuperGiant (RSG) and Asymptotic Giant Branch (AGB) stars. This choice is arbitrary but supported by the fact that these objects share similar observational techniques, observational properties, and main astrophysical questions.


\begin{figure}[!h]
\centering
\includegraphics[angle=0,width=0.9\hsize]{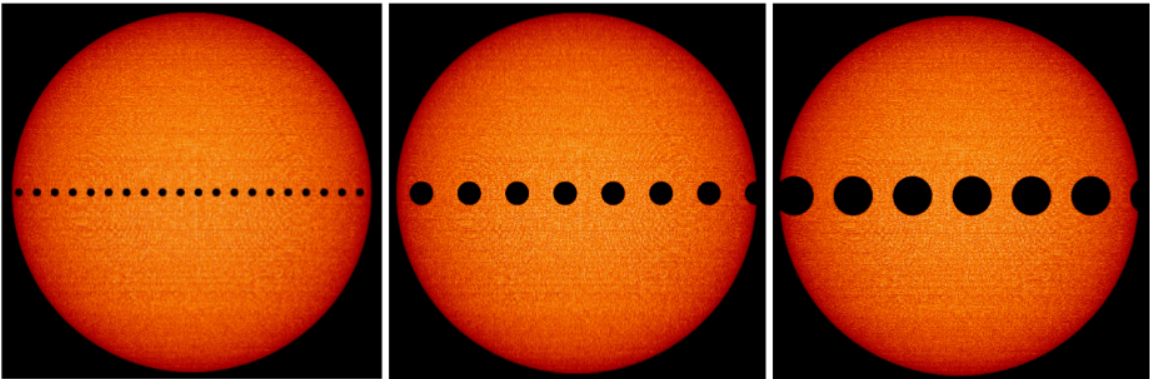} 
\caption[Example of a synthetic solar type star]{From main sequence stars to RGB phases. Example of a synthetic solar type star with terrestrial, neptunian, and hot Jupiter transiting planet.}
\end{figure}

\marginnote{\footnotesize \it From main sequence stars to RGB phases}Stars are not smooth. Their photosphere is covered by a granulation
pattern associated with the heat transport by convection. The
convection-related surface structures have different size, depth, and
temporal variations with respect to the stellar type. The related
activity (in addition to other phenomena such as magnetic spots,
rotation, dust, etc.) potentially causes bias in stellar parameters
determination, radial velocity, chemical abundances determinations,
and exoplanet transit detections. Convection manifests in the surface layers as a particular pattern of downflowing cooler plasma and bright areas where hot plasma rises \citep{2009LRSP....6....2N}. The size of granules depends on the stellar parameters of the star and, as a consequence, to the extension of their atmosphere \citep{2013arXiv1302.2621M}. Eventually, the convection causes inhomogeneous stellar surface that changes with time. They affect the atmospheric stratification in the region where the flux forms and the emergent spectral energy distribution, with potential effects on precise determinations of stellar parameters \citep{2011A&A...534L...3B,2012A&A...545A..17C, 2012A&A...540A...5C}, radial velocity \citep{2008sf2a.conf....3B,2011JPhCS.328a2012C,2013A&A...550A.103A}, chemical abundance \citep{2005ASPC..336...25A,2009ARA&A..47..481A,2011SoPh..268..255C}, photometric colours \citep{2018A&A...611A..11C,2017MmSAI..88...90B} and on planet detection \citep{2015A&A...573A..90M,2017A&A...597A..94C}. 

Convection is a difficult process to understand because it is nonlocal, and three-dimensional, and it involves nonlinear interactions over many disparate length scales. In this context, the use of numerical  three-dimensional radiative hydrodynamical simulations of stellar convection is extremely important. In recent years, with increased computational power, it has been possible to compute grid of 3D simulations (see Stagger code in Section~\ref{hydro}) that cover a substantial portion of the Hertzsprung-Russell diagram \citep{2013arXiv1302.2621M,2013ApJ...769...18T,2013A&A...558A..48B,2009MmSAI..80..711L}. With these tools, it is possible to predict reliable synthetic spectra and images for several stellar types. For a complete description of a star (and the planet hosted), a number of parameters are important, such as mass, luminosity, radius, age, pulsation period, chemical composition, angular momentum, magnetic field, mass-loss rate, and the circumstellar environment. The observational determination of several of these parameters requires a high spectral resolution. Furthermore, while some of these parameters are integrated quantities, several of them vary across the stellar disk and its environment, leading to the need for spatially resolved observations \citep{2004astro.ph.12519W}. In this context, the use of realistic 3D hydrodynamical stellar atmosphere simulations are essential for a large application range in addition to stellar convection studies and atmospheres themselves: stellar fundamental parameter determination, stellar spectroscopy and abundance analysis, asteroseismology, calibration of stellar evolutionary models, interferometry, and detection/characterisation of extrasolar planets.\\

\begin{figure}[!h]
\centering
\includegraphics[angle=0,width=0.8\hsize]{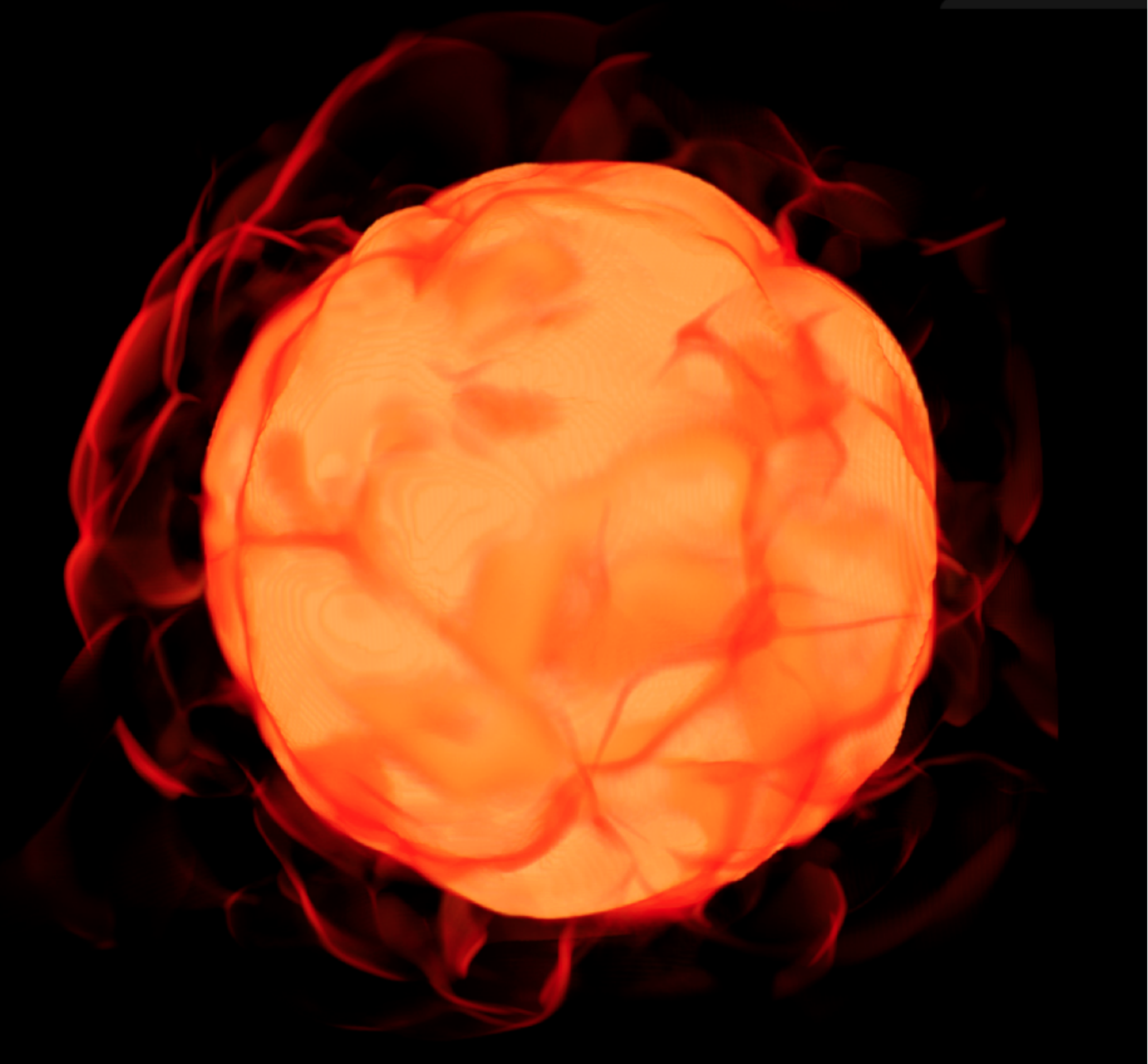} 
\caption[Example of evolved objects]{Evolved objects, synthetic RSG and AGB stars}
\end{figure}

\marginnote{\footnotesize \it Evolved objects, RSG and AGB stars}
Evolved cool stars of various masses are major cosmic engines \citep{2012ARA&A..50..107L}, providing strong mechanical and radiative feedback on their host environment. Through strong stellar winds and supernova ejections, they enrich their environment with chemical elements, which are the building blocks of the planets and life. In particular, these objects are known to propel strong stellar winds that carry the mass and angular momentum of the star's surface at speeds that vary with stellar brightness, evolution phase and chemical composition. In addition, stellar evolution models are not able to reproduce these winds without "ad hoc" physics. A complete understanding of their evolution in the near and distant Universe cannot therefore be achieved without a detailed understanding of wind physics across the life cycle of these stars as well as in relation to their cosmic environment. \\
The evolved cool stars can be divided in two categories based on their mass:

\begin{itemize}
\item Stars of low to intermediate mass ($0.5< M< 8M_\odot$, the exact value of the upper limit depends on the treatment of convection) evolve towards the red giant branch and the asymptotic branch of giants (AGB), increasing mass loss during this evolution. AGBs are characterized by very low effective temperatures (about 3000 K or less) and high luminosity (about 100 - 1000 L$_\odot$) which correspond to stellar radii of several hundred solar radii. Their high rate of mass loss  \citep[$\dot{M}$=10$^{-6}$-10$^{-4}$ M$_\odot$/yr, ][]{2010A&A...523A..18D} results from an interaction between pulsation, dust formation in the extended atmosphere and radiation pressure on dust \citep{2018A&ARv..26....1H}. This loss of mass is accompanied by a structural modification of their circumstellar envelopes, particularly pronounced during the evolution AGB $=>$ post-AGB $=>$ Planetary Nebula. The magnetic field coupled with the atmospheric dynamics should play a key role  in the process of initiating and maintaining a high mass loss \citep{2014A&A...561A..85L}, or together with the evolution of the geometry of the stellar wind.
\item More massive stars ($M > 8M_\odot$) evolve into red supergiant phase before they explode into type II supernovae. These stars are characterized by high luminosity (L $>$ 1000 L$_\odot$), effective temperatures between 3450 and 4100 K and stellar radii up to several hundreds of R$_\odot$, or even more than 1000 R$_\odot$ \citep{2005ApJ...628..973L}. One of the properties of RSG stars is their rate of mass loss \citep[$\dot{M}$=10$^{-7}$-10$^{-4}$ M$_\odot$/yr, ][]{1992ApJ...397..552W}, the origin of which being still poorly understood \citep[radiation pressure, radiative and/or pressure shock waves, magnetism, dustÉ, e.g., ][]{2007A&A...469..671J}. RSGs have a low intensity magnetic field (of the order of 1 gauss) that has been identified and monitored over several years \citep{2010A&A...516L...2A,2017A&A...603A.129T,2018A&A...615A.116M}. In these circumstances, the photospheric convective motions would tend to generate a local dynamo responsible for such intermittent fields \citep{2013LNP...857..231P}.
\end{itemize}

These stars are among the largest stars in the Universe and their luminosities place them among the brightest stars, visible to very long distances. This last point is very important for the quantitative analysis studies of metallicity in our Galaxy \citep[e.g., with JWST][]{2018arXiv181004187L} and in nearby galaxies \citep{2017ApJ...847..112D}. However, their chemical composition is very difficult to be obtained due to their complex spectra with broad, asymmetric lines with variations suspected to stem from a convection pattern  consisting of large granules and (super)-sonic velocities. 

The use of state-of-the-art numerical simulation of stellar convection (see CO$^5$BOLD code in Section~\ref{hydro}) is essential for the subsequent interpretation of the observations (spectroscopic, imaging, interferometry, astrometry...) and the solutions of the different open questions reported above. Interferometric observations, corroborated by the numerical simulations, show that the stellar surface of theses stars is covered by a few convective cells \citep{2010A&A...515A..12C,2010A&A...511A..51C,2017A&A...600A.137F,2017A&A...605A.108M} evolving on timescales of several weeks to years \citep{2018arXiv180802548C,2011A&A...528A.120C}, and causing large temperature inhomogeneities. Their velocity can levitate the gas and contribute to the mass loss \citep{2011A&A...535A..22C,2017A&A...600A.137F,2018A&A...610A..29K}.



\newpage
\section{Hydrodynamic modelling of stellar atmospheres}\label{hydro}

\lettrine[lines=3,nindent=4pt]
{L}{ate} type stellar atmosphere modelling is carried out using three-dimensional (3D) radiative hydrodynamical (RHD) simulations. The RHD codes solve the time-dependent, three-dimensional, compressible equations of radiative hydrodynamics in a cartesian grid. The horizontal dimensions are chosen to take into account several granules (typically 10 to 20) to extract a statistical behaviour. The depth of the simulation is adjusted so that at the base of the convective domain, the movements are homogeneous and adiabatic.

\begin{figure}[!h]
   \centering
  \begin{tabular}{cc}
  \includegraphics[width=0.35\hsize]{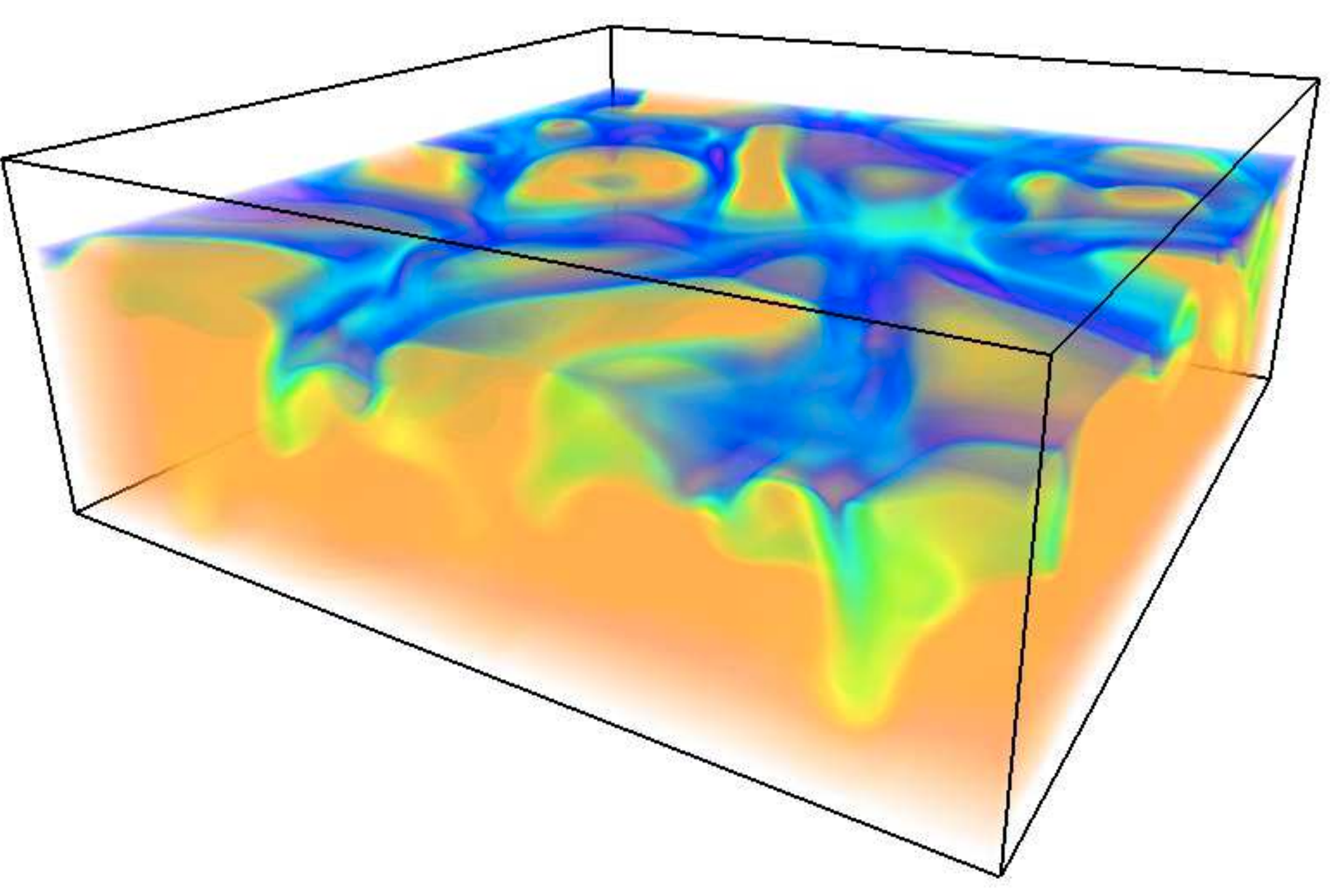}
  \includegraphics[width=0.35\hsize]{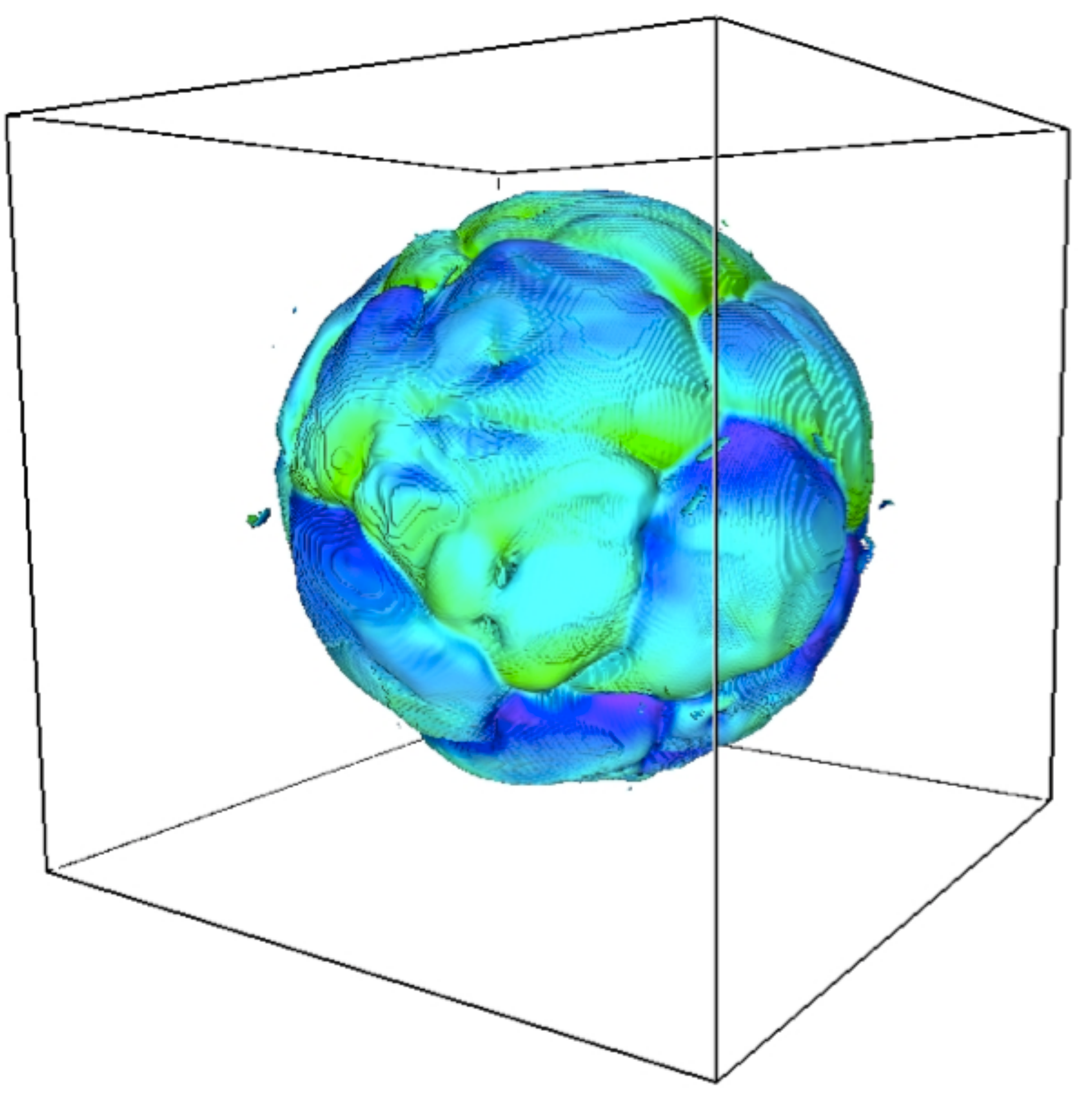} \\
 \end{tabular}
\caption[3D RHD simulations]{Setup of box-in-a-star (\emph{left}) and star-in-a-box (\emph{right}) configurations for 3D RHD simulations.}
              \label{fig:schema}%
\end{figure}

Two different geometries (Fig.~\ref{fig:schema}) can be used with 3D simulations:
\begin{itemize}
\item The \emph{box-in-a-star} configuration, where only a small portion of the upper layers is simulated. The computational time ranging from few days to few weeks depending
on the stellar type) cover only a small section of the surface layers
of the deep convection zone (typically ten pressure scale heights
vertically), and the numerical box includes about $\sim10$ convective
cells, which are large enough so that the cells are not constrained by
the horizontal (cyclic) boundaries, while the bottom and upper boundaries are open. A constant gravitation is considered.
\item The \emph{star-in-a-box} configuration, where the entire convective envelope is taken into account. In fact, when the stellar surface gravity is lower than $\log g \sim 1$, the box-in-a-star
simulations become inadequate because of the influence of sphericity becomes important. The computational domain is a cubic grid equidistant in all directions, and the same open boundary condition is employed for all sides of the computational box. A fixed external spherically symmetric gravitational field is used. The simulations computed with this setup are highly computer-time demanding and difficult to
run, which is the reason why there are only few (less than $\approx$30) simulations available so far.

\end{itemize}

The \emph{box-in-a-star} setup can be used as long as the granule size is small compared to the size of
the star and the sphericity effects are negligible. It is mandatory to use a \emph{star-in-a-box} approach for very low surface gravity stars, in particular RSG and AGB stars, where the granule size becomes extremely large. For all other stars in the main sequence up to RGB phase, it is fair to use the \emph{box-in-a-star} configuration. \\ 
\marginnote{\footnotesize \it No fudge parameters in 3D RHD simulations}Unlike conventional 1D atmospheric models, which require the introduction of arbitrary quantities (e.g., micro and macro-turbulence) to straw their lack of physical realism, RHD simulations are ab-initio and without free parameters. The crucial point of the RHD codes for stellar atmospheres lies in the coupling between hydrodynamic equations and radiative transfer with realistic opacities and adapted equations of state. A detailed and precise solution of the radiative transfer is essential for a realistic treatment of convection because it is the radiative losses in the surface layers of the star that drain the convective movements and thus influence the whole simulation domain.

The RHD codes I use are:

\begin{itemize}
\item The CO$^5$BOLD code \citep{2012JCoPh.231..919F} for the RHD simulations in the \emph{star-in-a-box} configuration (i.e., global simulations). 
\item The STAGGER code \citep[Nordlund et Galsgaard 1995\footnote{http://www.astro.ku.dk/$\sim$kg/Papers/MHD\_code.ps.gz},][]{2009LRSP....6....2N,2011A&A...528A..32C} for the RHD simulations in the \emph{box-in-a-star} configuration (i.e., local simulations).
\end{itemize}

\marginnote{\footnotesize \it CO$^5$BOLD code}The CO$^5$BOLD code solves the coupled non-linear equations of compressible hydrodynamics (with an approximate Roe solver, \cite{}) and non-local radiative energy transfer (for global simulations with a short-characteristics scheme) in the presence of a fixed external gravitational field and in a 3D cartesian grid. The equation of state uses pre-tabulated values as functions
of density and internal energy $\left(\rho,e_i\rightarrow P,\Gamma_1,T,s\right)$. It accounts for H{\sc{I}}, H{\sc{II}}, H$_{2}$, He{\sc{I}}, He{\sc{II}}, He{\sc{III}} and a representative metal for any prescribed chemical composition. The equation of state
does not account for the ionization states of metals, but it uses
only one neutral element to achieve the appropriate atomic weight
(in the neutral case) for a given composition. 
The radiation transport step for global simulations uses a short-characteristics
method and the frequency dependance of the radiation field is generally the gray approximation, which completely ignores the frequency dependence, for computational time reasons. However, for few global simulations the radiative transfer is calculated for up to five wavelengths groups according to the run of the monochromatic optical depth in
a corresponding MARCS \citep{2008A&A...486..951G} 1D model.

The code can be used for both local and global setups. Motions in the core are damped by a drag force to suppress dipolar oscillations. The hydrodynamics and the radiation transport scheme ignore the core completely and integrate right through it. The code is parallelized with hybrid Open Multi-Processing (OpenMP) and MPI directives. \\
For more details, see \cite{2011A&A...535A..22C} for the computation of RSG simulations and \cite{2017A&A...600A.137F} for the AGB ones.


\marginnote{\footnotesize \it Stagger code}In the Stagger-code, the equations for the conservation of mass, momentum, and energy for compressible flows together with the induction equation are discretised using a high- order finite-difference scheme (sixth-order numerical derivatives, fifth-order interpolations) and solved on a rectangular Eulerian mesh, extending over a representative 3D, rectangular, volume located across the optical surface and including the photospheric layers as well as the upper part of the stellar convection zone. The simulation domains are chosen large enough to cover at least ten pressure scale heights vertically and to allow for about ten granules to develop at the surface. At the bottom of the simulation, the inflows have a constant entropy, and the whole bottom boundary is set to be a pressure node for p-mode oscillations. The code employs realistic input physics: the equation of state is an updated version of the one described by \cite{1988ApJ...331..815M}, and the radiative transfer is calculated for a large number over wavelength points merged into 12 opacity bins \citep{1982A&A...107....1N,2000ApJ...536..465S,2013A&A...557A..26M}. They include continuous absorption opacities and scattering coefficients from \cite{2010A&A...517A..49H} as well as line opacities described in \cite{2008A&A...486..951G}, which in turn are based on the VALD-2 database \citep{2001ASPC..223..878S} of atomic lines. The code can be used for local setup only and is parallelized with MPI directives. 

\textit{\textcolor{blue}{The paper, "Radiative hydrodynamics simulations of red supergiant stars. IV gray versus non-gray opacities" by Chiavassa et al. 2011, is attached at the end of the chapter.}}.

\section{Post-processing multidimensional radiation transport}

\lettrine[lines=3,nindent=4pt]
{D}{uring} my Ph.D and afterwards, I developed a pure LTE radiative transfer code, {{\sc Optim3D}} to generate synthetic spectra and intensity maps from snapshots of the 3D RHD simulations, taking account the Doppler shifts caused by the convective motions \citep{2009A&A...506.1351C}. The radiation transfer is calculated in detail using pre-tabulated extinction coefficients based on MARCS opacities for molecules \citep{2008A&A...486..951G} and VALD database \citep{1995A&AS..112..525P} for atomic lines. These opacities are generated using TURBOSPECTRUM code \citep{1998A&A...330.1109A,2012ascl.soft05004P}. These tables are functions of temperature, density and wavelength, and are computed with the preferred solar (or not-solar) composition. For RHD simulations, no micro- and macro-turbulence is considered. The temperature and density distribution is optimised to cover the values encountered in the outer layers of the simulations. Eventually, the wavelength resolution is generally at least 3 times larger than the expected output and, at least, R = $\lambda/\Delta\lambda\geq500\ 000$. 

\marginnote{\footnotesize \it{In practice, what {{\sc Optim3D}} does}}In practice, 3D simulations provide a thermodynamic structure of the outer layers of the atmosphere (i.e., temperature, density and velocity field, as a function of optical depth). This structure is read by {{\sc Optim3D}}. The latter then calculates the monochromatic intensity emerging with respect to the observer along a line-of-sight perpendicular to a face of the cube, and/or along several inclined rays. {{\sc Optim3D}} interpolates the opacity tables in temperature and logarithmic density for all the simulation grid points using a bi-linear interpolation. The interpolation coefficients are computed only once, and stored. 
Bi-linear interpolation has been preferred to spline interpolation because: (i) spline is significantly more time consuming, 
(ii) and comparisons with other codes do not show noticeable improvements using splines. Then, the logarithmic extinction coefficient is linearly interpolated at each Doppler-shifted wavelength in each cell along the ray, 
and  the optical depth scale along the ray is calculated. The numerical integration of the equation of radiative transfer gives the intensity emerging towards the observer at that wavelength and position. This calculation is performed for every line-of-sight perpendicular to the face of the computational box, and for all the required wavelengths.

{{\sc Optim3D}} was initially conceived to be used with CO$^5$BOLD \emph{star-in-a-box} simulations \citep{2009A&A...506.1351C}. In \cite{2010A&A...524A..93C}, I extended the code to \emph{box-in-a-star} configuration computed with Stagger-code. I implemented a procedure to tilt the computational box (Fig.~\ref{fig:schema}) by an angle $\theta$ with respect to
the line of sight (vertical axis) and rotating it azimuthally by an angle $\phi$. The final result is a spatially resolved intensity spectrum at different angles. More recently, Kateryna Kravchenko implemented the computation of the contribution function to the line depression in {{\sc Optim3D}} aiming at correctly identifying the depth of formation of spectral lines in order to construct numerical masks probing spectral lines forming at different optical depths \citep{2018A&A...610A..29K}. {{\sc Optim3D}} is optimised to be fast (slighter slower than a 1D calculation) also with high resolution simulations, and it can be easily parallelised on a cluster of computers.

\begin{figure}[!h]
   \centering
  \begin{tabular}{c}
  \includegraphics[width=1.0\hsize]{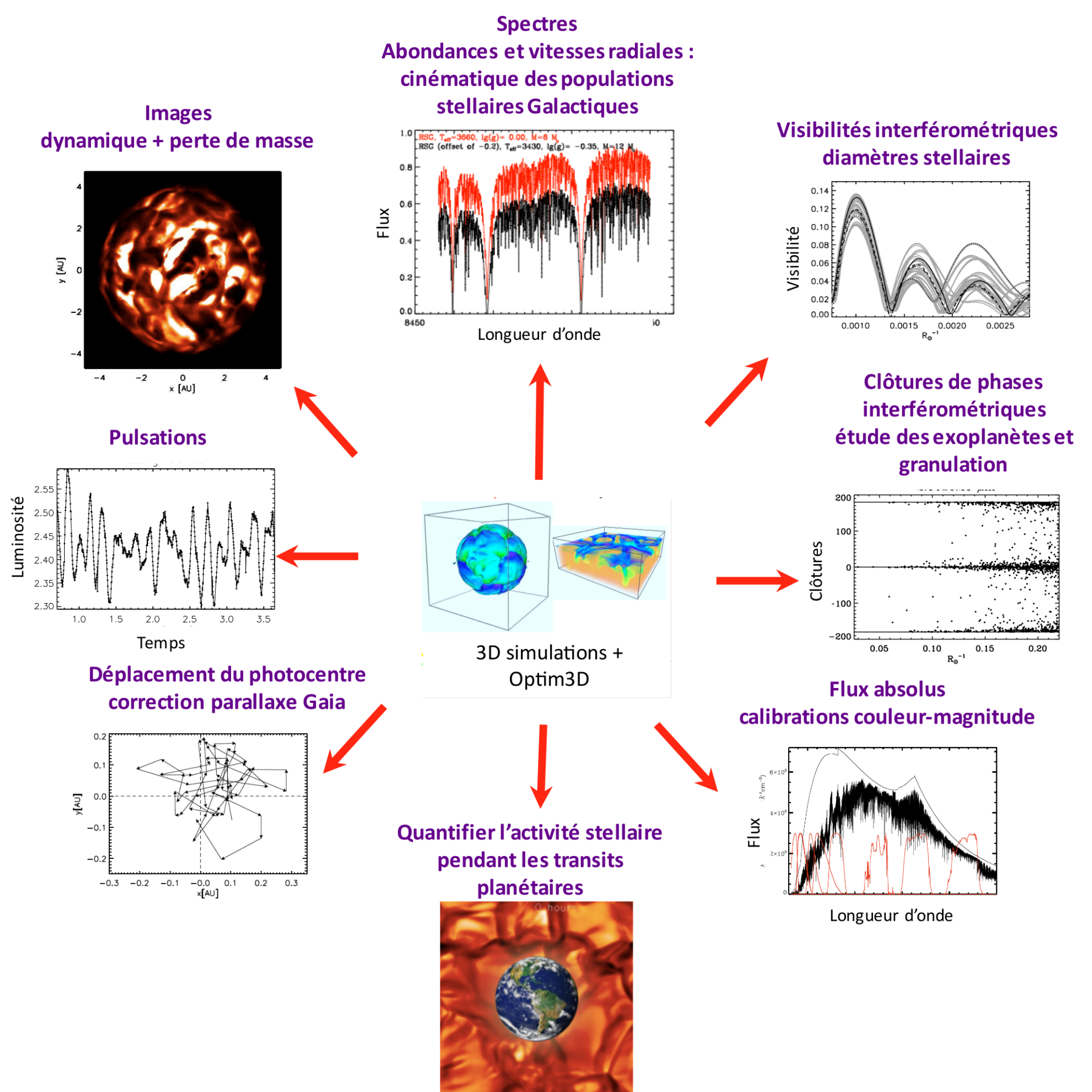}
  \end{tabular}
\caption[{{\sc Optim3D}} post-processing code]{Scientific problems tackled thanks to the synergy between the 3D RHD simulations and the detailed calculations of the post-processing code {{\sc Optim3D}}}
              \label{schema}%
\end{figure}

The synergy between 3D RHD simulations and the detailed calculation of the radiative transfer with {{\sc Optim3D}} makes it possible to approach a broad spectrum of astrophysical problems which are summarised in Fig.~\ref{schema}. \\

\textit{\textcolor{blue}{The paper, "Radiative hydrodynamics simulations of red supergiant stars: I. interpretation of interferometric observations" by Chiavassa et al. 2009, is attached at the end of the chapter.}}.

\invisiblesection{\it Attached paper: \textbf{RHD simulations}, Red Supegiants and Co5Bold. \cite{2011A&A...535A..22C}}
\includepdf[pages={1-14}]{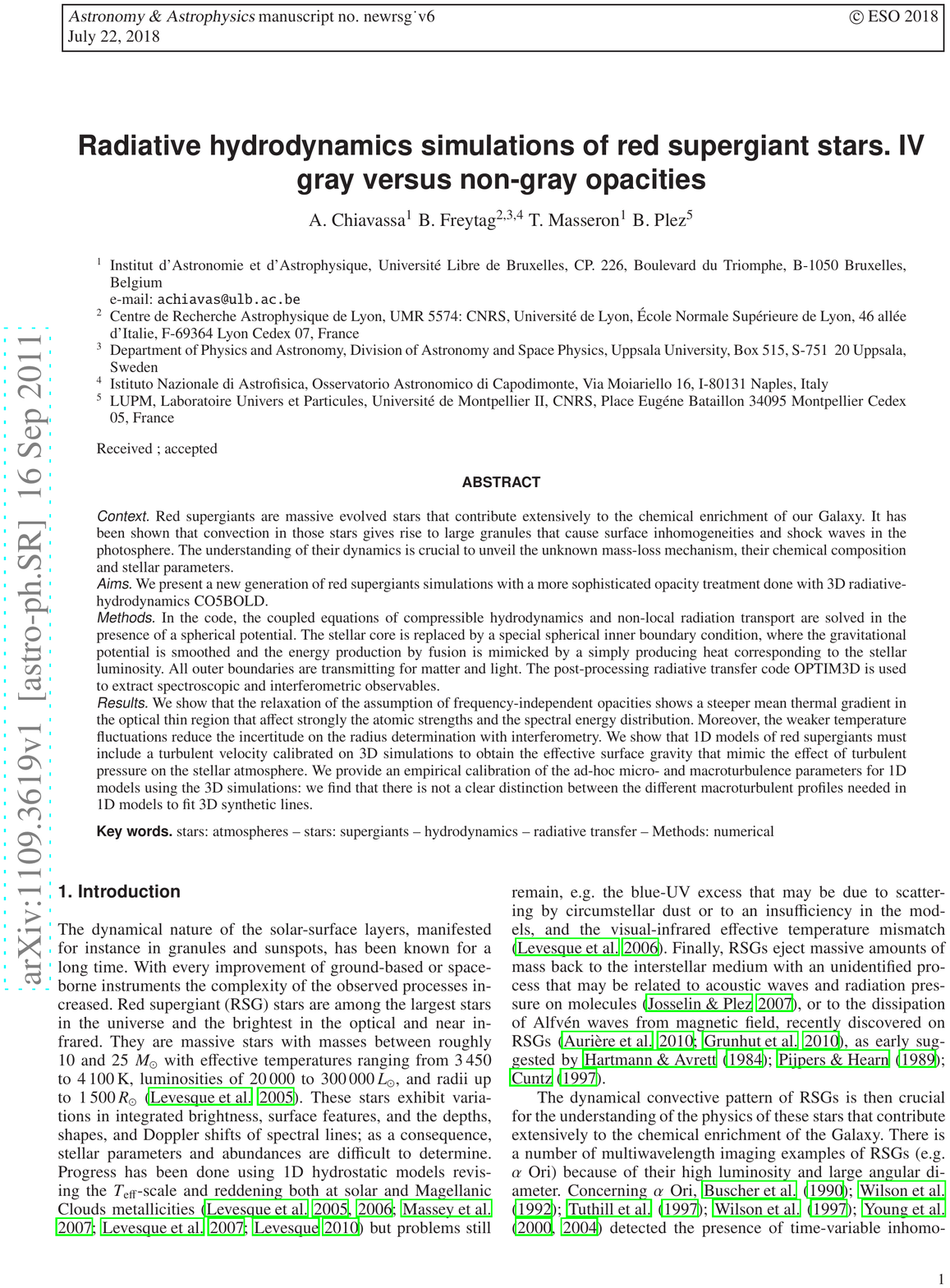}
\invisiblesection{\it Attached paper, \textbf{Optim3D} and Red Supegiants. \cite{2009A&A...506.1351C}}
\includepdf[pages={1-15}]{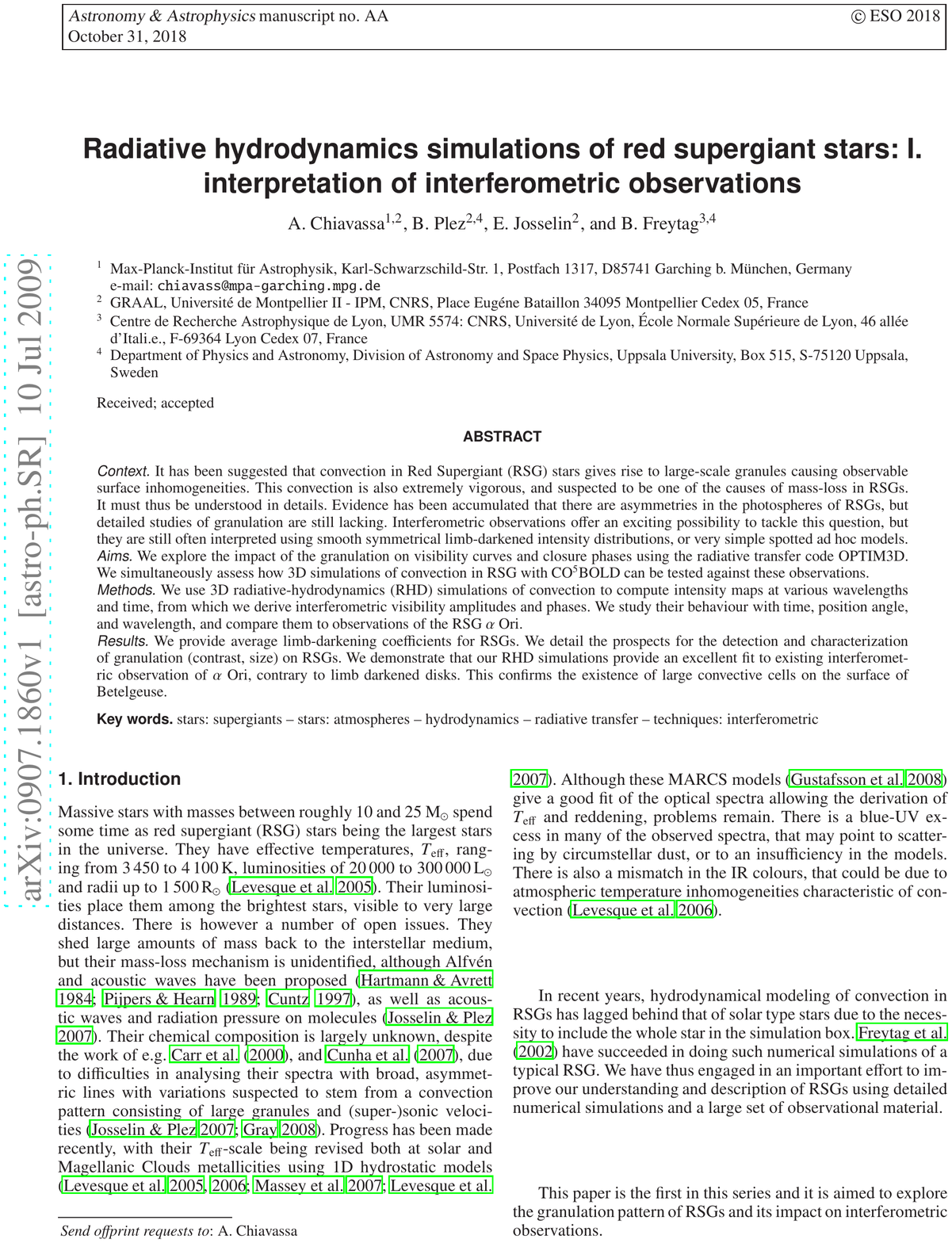}
\clearemptydoublepage

\chapter{A journey cross Hertzsprung-Russel diagram \label{ch:intro2}}

\begin{figure}[!h]
   \centering
  \begin{tabular}{cc}
  \includegraphics[width=0.4\hsize]{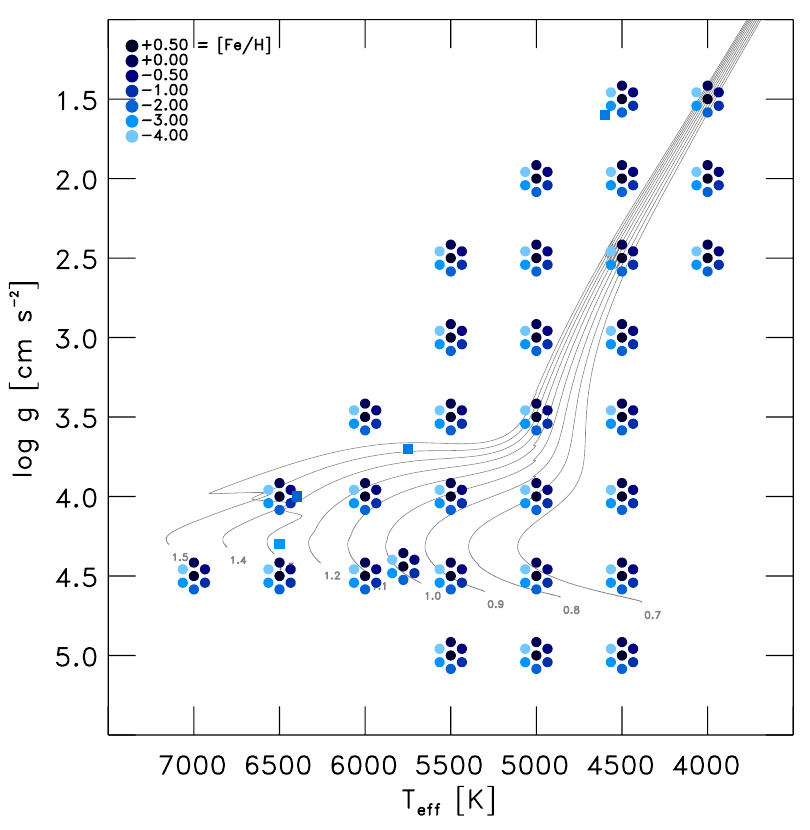}
  \includegraphics[width=0.4\hsize]{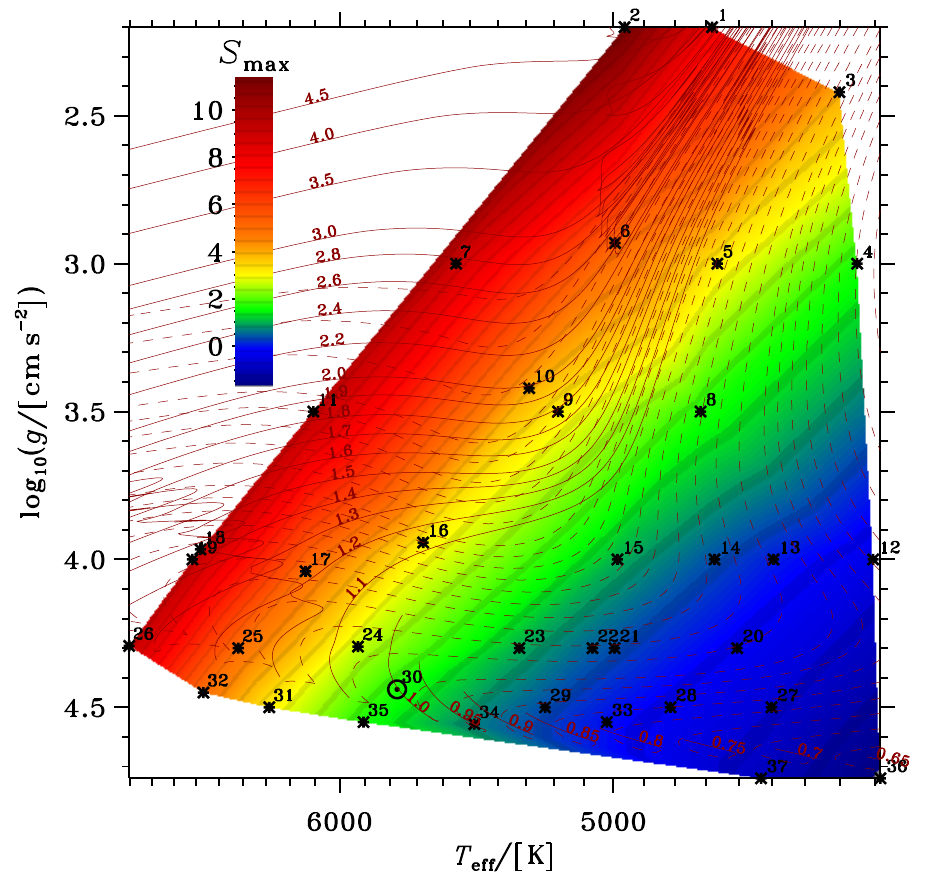}\\
  \includegraphics[width=0.4\hsize]{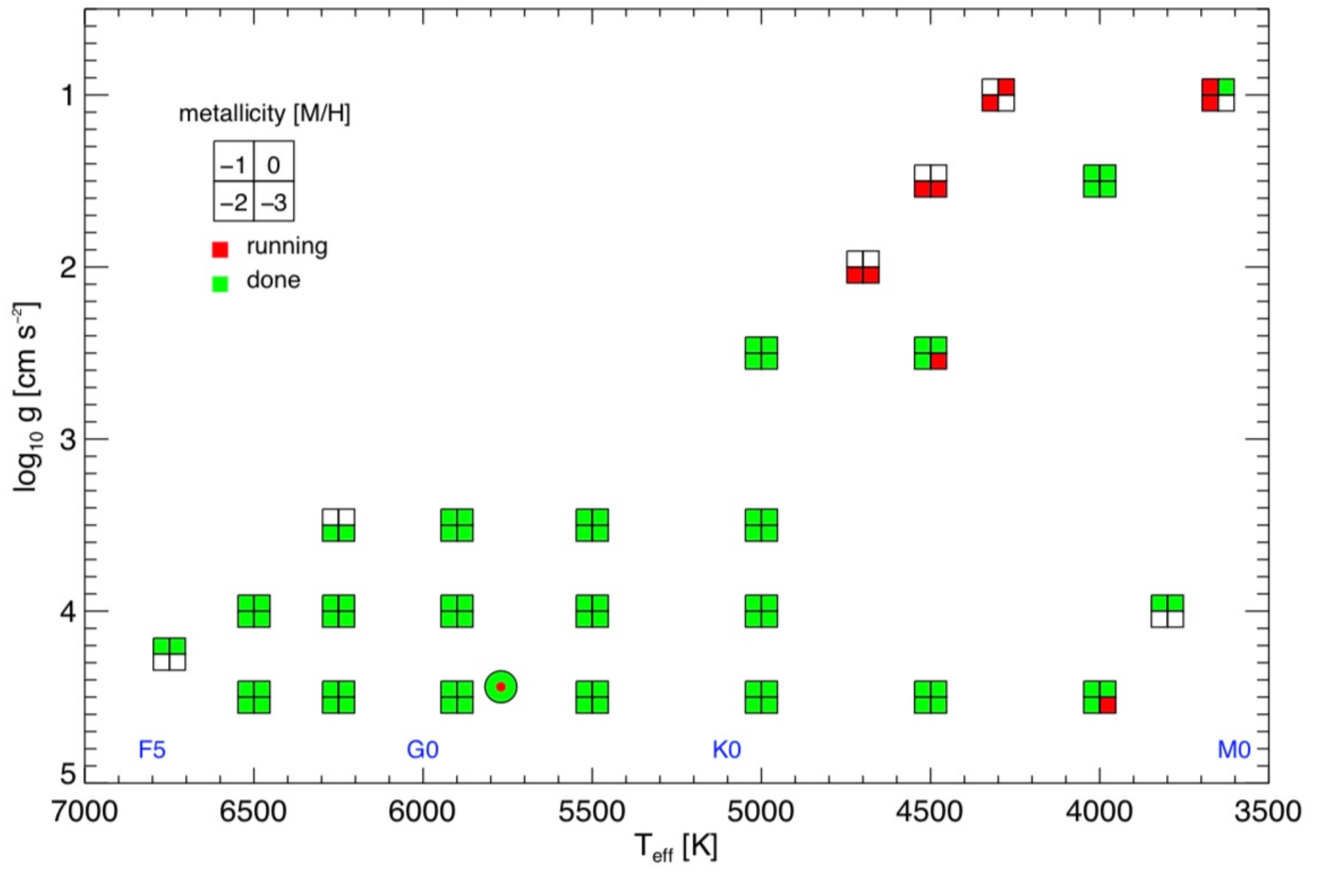}
  \includegraphics[width=0.4\hsize]{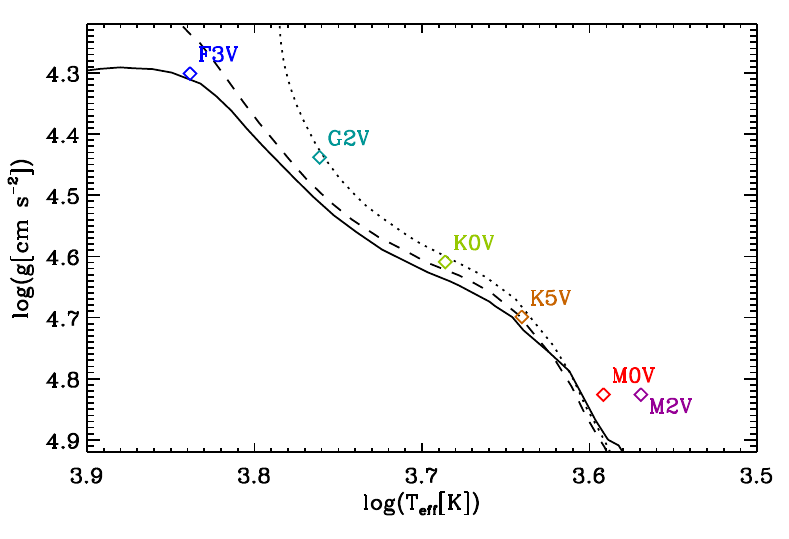}\\
   \includegraphics[width=0.37\hsize]{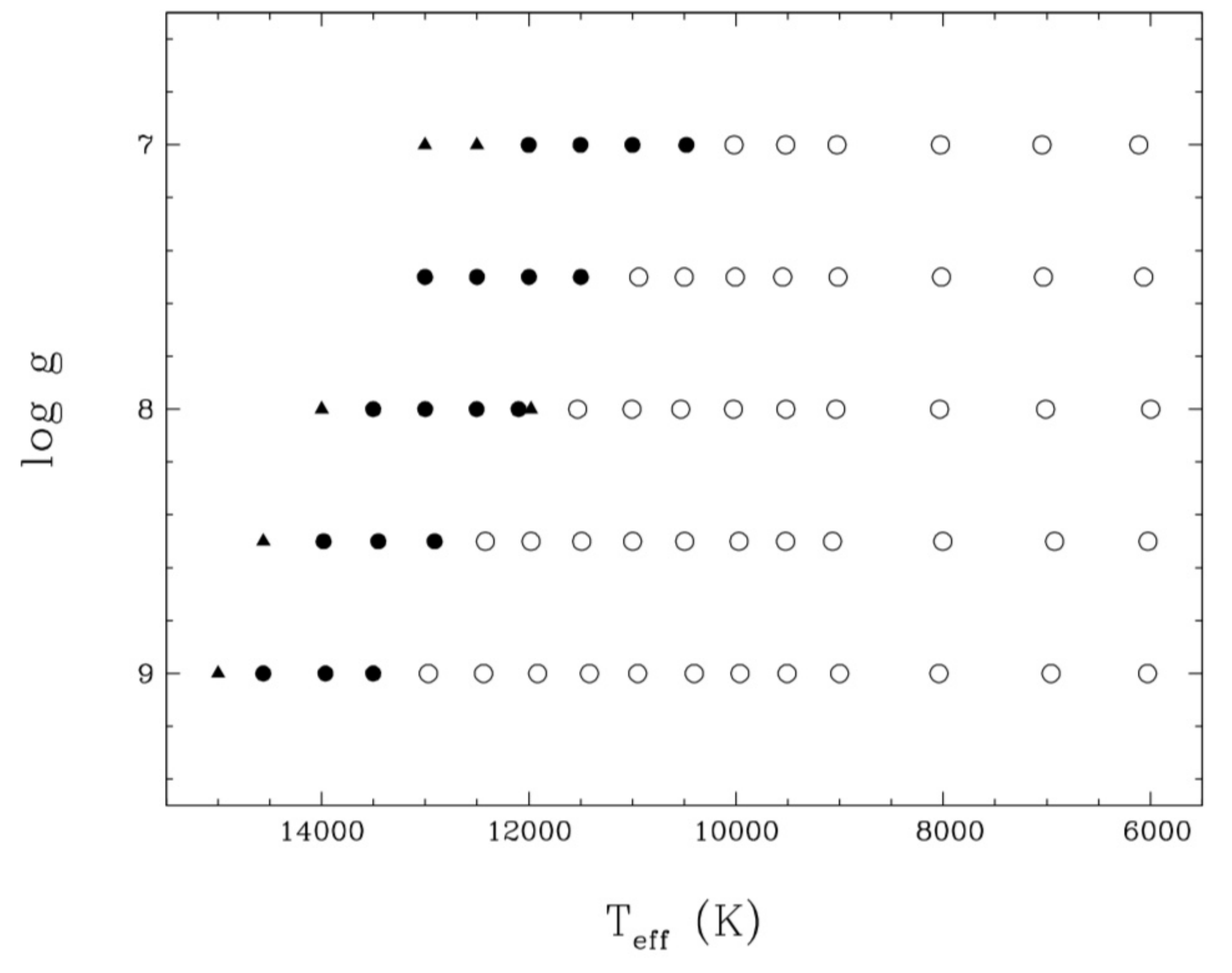}
  \includegraphics[width=0.4\hsize]{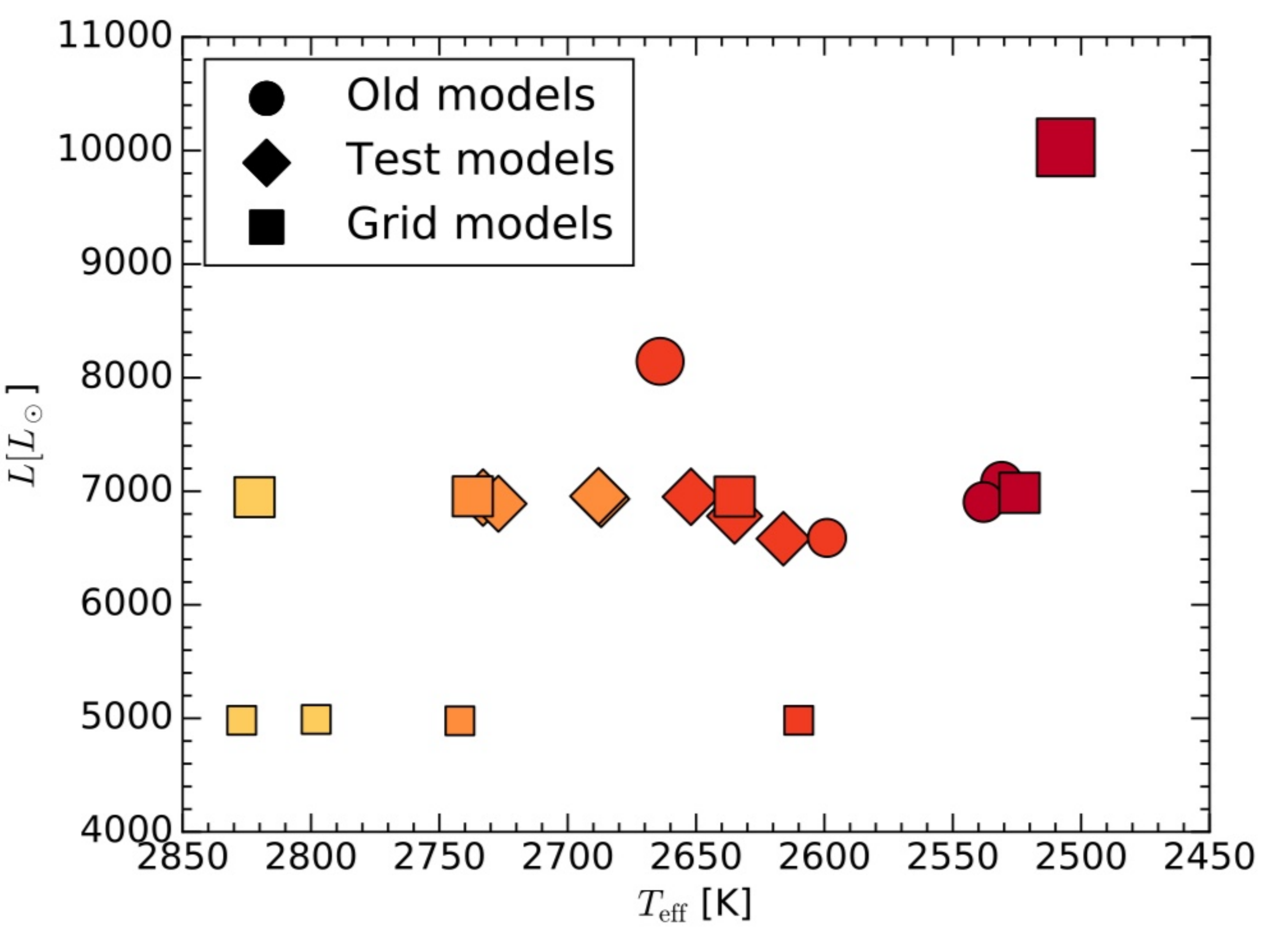}\\
 \end{tabular}
\caption[3D RHD simulation-grid in the H-R diagram]{3D RHD simulation-grid in the H-R diagram. \emph{Top panels:} simulations computed with Stagger-code or similar branches (Section~\ref{hydro}), Stagger-grid \citep[left,][]{2013A&A...557A..26M} and "Trampedach" grid \citep[right,][]{2013ApJ...769...18T}. \emph{Central panels:} the CIFIST grid (left, \citep{2009MmSAI..80..711L} computed with CO$^5$BOLD code, and the grid \citep[right,][]{2013A&A...558A..48B} computed with MURaM core \citep{2005A&A...429..335V}. \emph{Bottom panels:} the White Dwarf grid \citep[left,][]{2013A&A...559A.104T} and the AGB one \citep[right,][]{2017A&A...600A.137F} computed with CO$^5$BOLD code (Section~\ref{hydro}).}
              \label{fig:3Dgrids}%
\end{figure}

\lettrine[lines=3,nindent=4pt]{T}{here} are a few hundred billion stars in our Galaxy, and billions of galaxies in the Universe. One important technique in science is to try and sort/classify things into groups and seek out trends or patterns.
This is done using stars. These can be plotted to form what is one of the most useful plots for stellar astronomy, the Hertzsprung-Russell (or H-R) diagram. Hertzsprung \citep{1905WisZP...3..442H} and later independently Russell \citep{1919PNAS....5..391R} realized that the knowledge of the absolute brightness of stars together with their spectral type or surface temperature allowed fundamentally different types of stars to be distinguished. Both astronomers already realized that the apparent stellar brightness was insufficient to draw conclusions, and that the absolute magnitude (i.e., distance) was a crucial need to properly order the stars in the H-R diagram. They also pointed out that the diagram contains important information about the stellar radii, with the giant sequence to be found at a larger brightness but similar surface temperatures to the cools stars of the dwarf or main sequence. Underlining a clear link with the qualitative determination of a stellar radius using the Stefan-Boltzmann law. \marginnote{\footnotesize \it The H-R diagram is an essential diagnostic diagram}The H-R diagram is an essential diagnostic diagram for stellar structure and evolution, which has now been in use for more than 100 years \citep{1964Cent....9..219N}.

With the advent of stellar model atmospheres, it has been possible to calculate accurately the surface gravity, e.g. \cite{1972ApJS...24..193A} , that allowed to order the stars in the surface gravity-effective temperature (i.e., Kiel diagram).
The main advantage of this digram is that stars can be compared to stellar evolution predictions without prior knowledge of their distances, e.g., a first example is in \cite{1976A&A....52...11K}. However, the comparison is often negatively affected by the relatively large uncertainties of the spectroscopic gravity determinations. In addition to this, the radius or any other stellar parameters cannot be directly retrieved from this approach and other technique are needed, e.g., interferometry \citep{2004A&A...426..297K} and asteroseismology \citep{2009MNRAS.400L..80S}.

\marginnote{\footnotesize \it 3D RHD simulations of cool stars across the H-R diagram}In recent years, with increased computational power, it has been possible to compute grids of 3D RHD simulations (Section~\ref{hydro}) covering largely the H-R diagram. Fig.~\ref{fig:3Dgrids} displays how the different codes available attack this problem. The computational time required for each simulation varies largely with stellar parameters, the numerical resolution, the physical ingredients included, the numerical approaches and the directive of parallelization taken into account. For a detailed comparison between the different codes, the reader may refer to \cite{2012A&A...539A.121B}. All these grids are constantly developing and some of the figures displayed may be out of date. Grids with limited number of simulations exist for particular stars: e.g., Cepheids stars \citep{2017A&A...606A.140V} , Red Supergiant stars \citep{2011A&A...535A..22C}, M-dwarf \citep{2013AN....334..137W}, Brown dwarf \citep{2013MSAIS..24..128A}.

\section{Main Sequence stars up to RGB phase\label{sec:mainseq}}

\lettrine[lines=3,nindent=4pt]{I }{start} the review of my results with main sequence stars. In 2013, I participated in the publication of the Stagger-grid \citep{2013A&A...557A..26M}. This grid is intended for various applications in addition to stellar convection studies and atmospheres themselves, including stellar parameter determination, stellar spectroscopy and abundance analysis, asteroseismology, calibration of stellar evolution models, interferometry, and characterisation of extrasolar planets. The surface structures and dynamics of cool stars are characterised by the presence of convective motions and turbulent flows which shape the emergent spectrum and the size of granules depends on the stellar parameters of the star and, as a consequence, on the extension of their atmosphere (Fig.~\ref{fig:3Dgran}).

\subsection{Stellar fundamental parameters with spectroscopy}

\begin{figure}[!h]
   \centering
  \begin{tabular}{cc}
  \includegraphics[width=0.52\hsize]{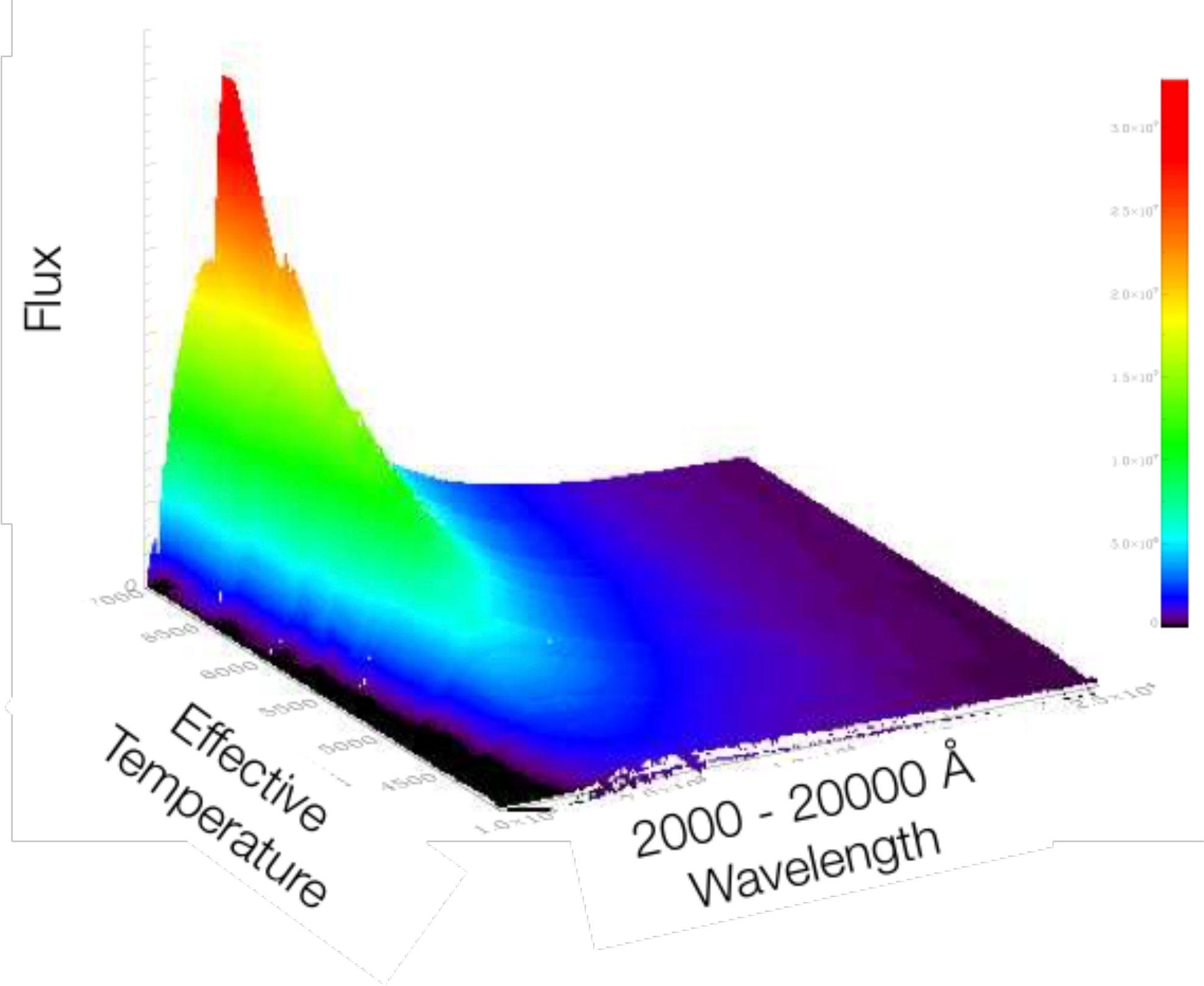}
  \includegraphics[width=0.52\hsize]{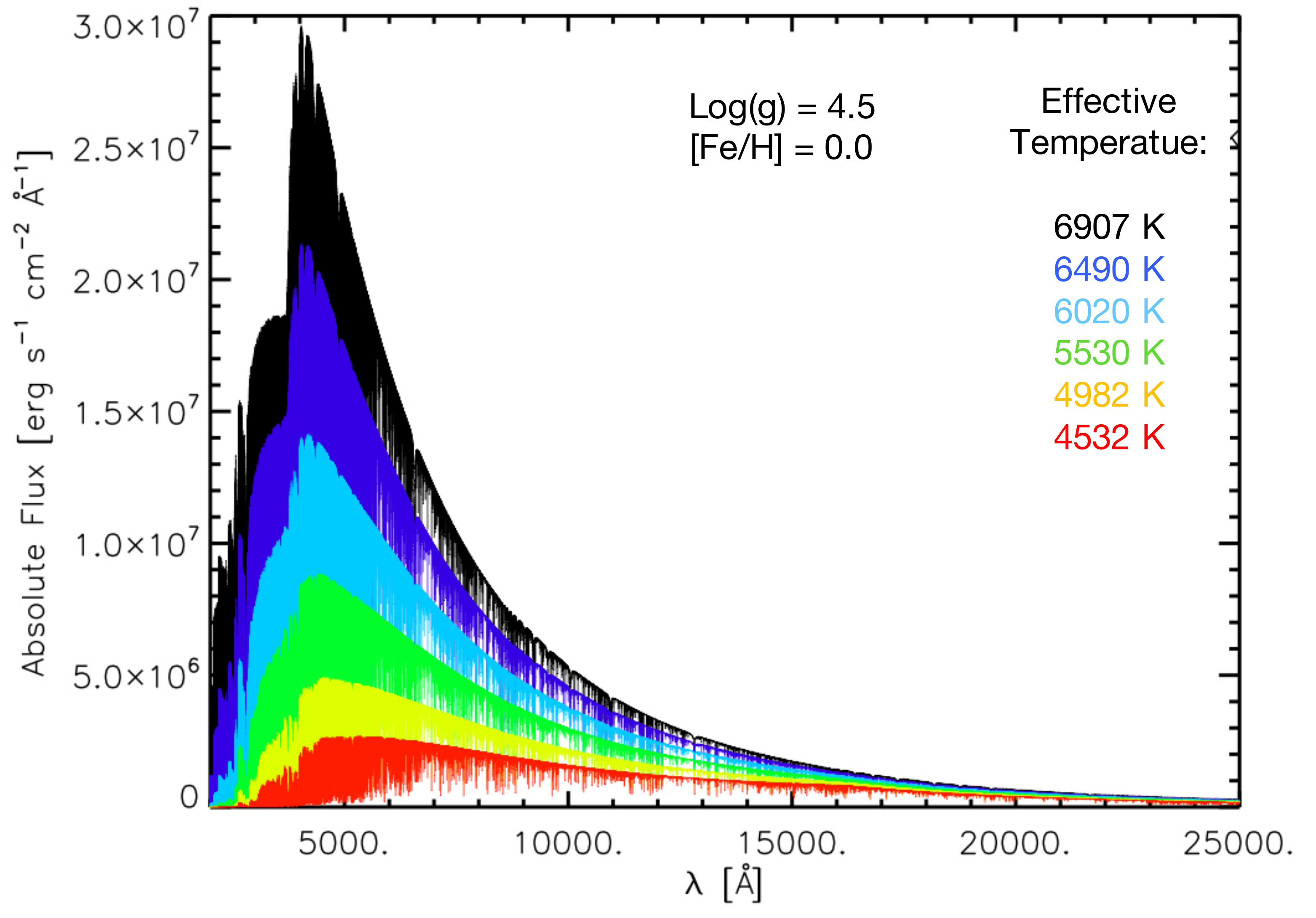}\\
  \includegraphics[width=0.4\hsize]{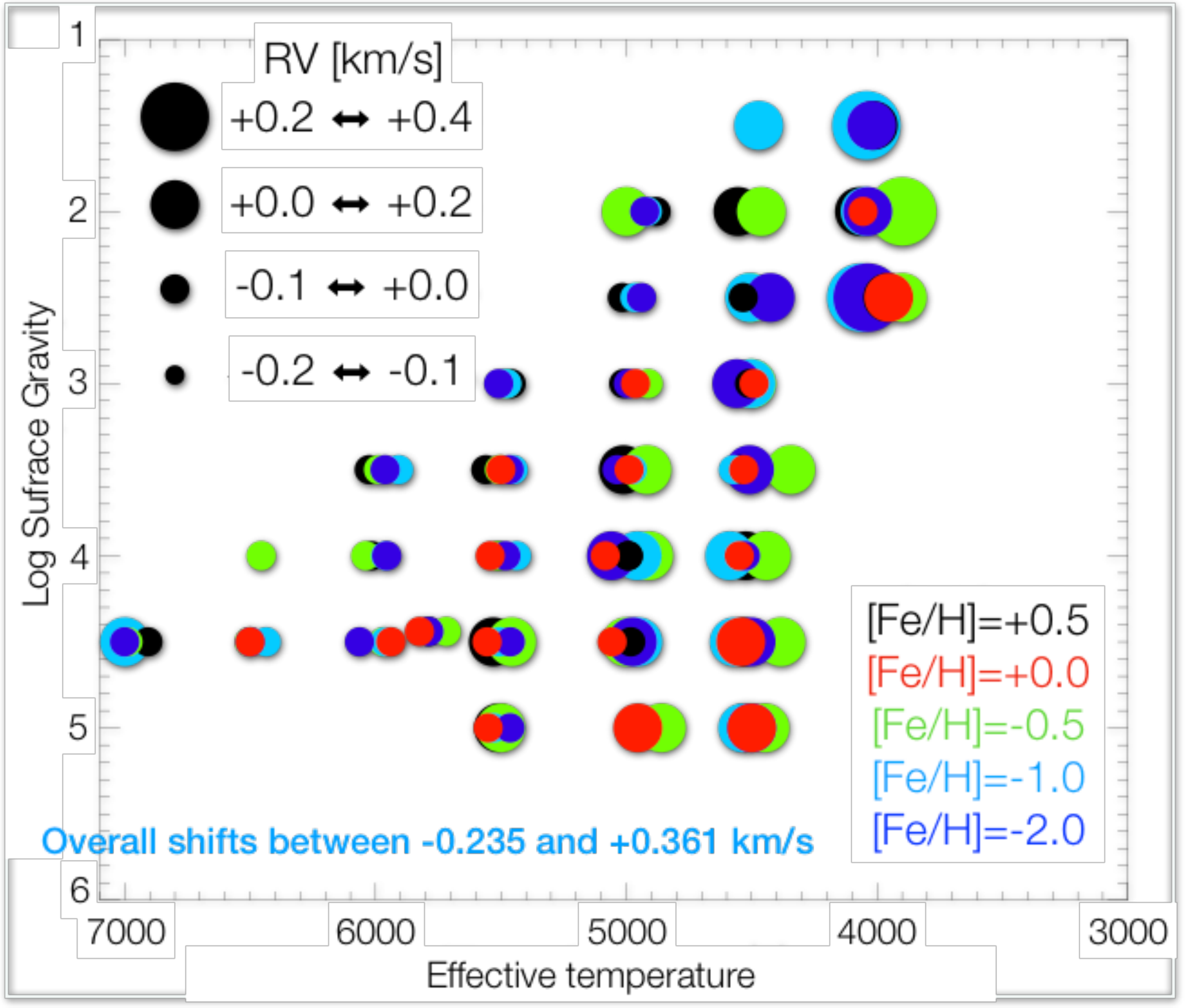}
 \end{tabular}
\caption[Synthetic spectra across the H-R diagram]{\emph{Top left:} Surface rendering for all the synthetic spectra computed for the 3D RHD simulations \citep{2018A&A...611A..11C}. For clarity, the wavelength  range has been reduced to 0.2 -- 2.5 $\mu$m. \emph{Top right:} Example of different spectra for RHD simulations with fixed surface gravity and metallicity, and varying effective temperature. \emph{Bottom left}: Convective shifts predicted by the 3D RHD simulations for the Gaia RVS spectral range and the Fe $\mathrm{I}$ lines.} 
              \label{fig:3Dflux}%
\end{figure}

A first application of this grid is the library of high-resolution stellar synthetic fluxes obtained from these 3D simulations \citep{2018A&A...611A..11C}. These spectra are calculated spectra from 1000 to 200 000 \AA\ with a constant resolving power of $\lambda$/$\Delta\lambda$=20\ 000 (Fig.~\ref{fig:3Dflux}, top panels) and from 8470 and 8710 \AA\ (Gaia Radial Velocity Spectrometer - RVS - spectral range) with a constant resolving power of $\lambda$/$\Delta\lambda$=300\ 000. I used the synthetic spectra to compute theoretical colours in the Johnson-Cousins UBV(RI)C, SDSS, 2MASS, Gaia, SkyMapper, Str\"omgren, and HST-WFC3 systems.

\marginnote{\footnotesize \it The 3D stellar synthetic grid} The synthetic magnitudes are compared with those obtained using 1D hydrostatic models. 1D versus 3D differences are limited to a small percent ($<5\%$) except for the narrow filters that span the optical and UV region of the spectrum (up to $\approx 10\%$). All the spectra publicly available on POLLUX database\footnote{http://pollux.graal.univ-montp2.fr}. 3D Bolomentric corrections are available on CDS database.

 \marginnote{\footnotesize \it Radial velocities}	Measurements of stellar radial velocities are fundamental in order to determine stellar spacevelocities. This is needed, for example to investigate the kinematic structure of stellar populations in the Galaxy or to monitor for radial velocity variations, either of which would point to the presence of unseen companion(s). Convection plays a crucial role in the formation of spectral lines and deeply influences the shape, shift, and asymmetries of lines in late-type stars \citep{2000A&A...359..669A}. Absorption lines may be blueshifted as a result of convective movements in the stellar atmosphere: bright and rising convective elements contribute  more photons than the cool dark shrinking gas, and as a consequence, the absorption lines appear blueshifted \citep{1982ARA&A..20...61D}. However, the convective line shift is not the same for all the spectral lines. Each line has a unique fingerprint in the spectrum that depends on line strength, depth, shift, width, and asymmetry across the granulation pattern depending on their height of formation and sensitivity to the atmospheric conditions. In \cite{2018A&A...611A..11C}, we determined the convective shift (cross-correlation of each 3D spectrum with the corresponding 1D) considering only few unblended Fe $\mathrm{I}$ and only Ca II triplet lines in the Gaia RVS range (8470 to 8710 \AA). The  values for the Fe $\mathrm{I}$ are in the range between  -0.235 and +0.361 km/s while the Ca II lines are strongly redshifted (Fig.~\ref{fig:3Dflux}, bottom left panel).

Despite the very recent publication in march 2018, the 3D spectral grid has been used in several works. I report here few examples:

\begin{itemize}
\item \cite{2018A&A...616A..39M} used the spectra from \cite{2015A&A...573A..90M} and \cite{2018A&A...611A..11C} to compute the power-2 limb-darkening for several passbands (UBVRI, CHEOPS, TESS, Kepler). He finds a very good agreement between observations (exoplanet systems and binary stars) and limb-darkening constructed on 3D intensity profiles. He proposes a new powerful analysis of light curves for transiting exoplanet systems and binary stars for stars with stellar parameters within the Stagger-grid range.
\item \cite{2018arXiv180406344Z} used the spectra to extract accurate radial velocities for GALAH DR2 data release. They achieved a typical accuracy of 0.1 km/s for about 212\ 000 stars in the well-populated regions of the HR diagram for stars with metallicity between -0.6 and +0.3. The level of accuracy achieved is adequate for studies of dynamics within stellar clusters, associations and streams in the Galaxy.
\item Gaia-DR3 release will provide accurate radial velocities from RVS. These values needs an appropriate wavelength calibration from convective shifts. This is directly processed in RVS pipeline using 3D synthetic spectra (Gaia consortium-CU6).
\end{itemize}

\textit{\textcolor{blue}{The paper, "The Stagger-grid: A grid of 3D stellar atmosphere models V. Synthetic stellar spectra and broad-band photometry" by Chiavassa et al. 2018, is attached at the end of the chapter.}}.

\subsection{Stellar fundamental parameters with interferometry}\label{sec:param}

\begin{figure}[!h]
   \centering
  \begin{tabular}{cc}
  \includegraphics[width=1.0\hsize]{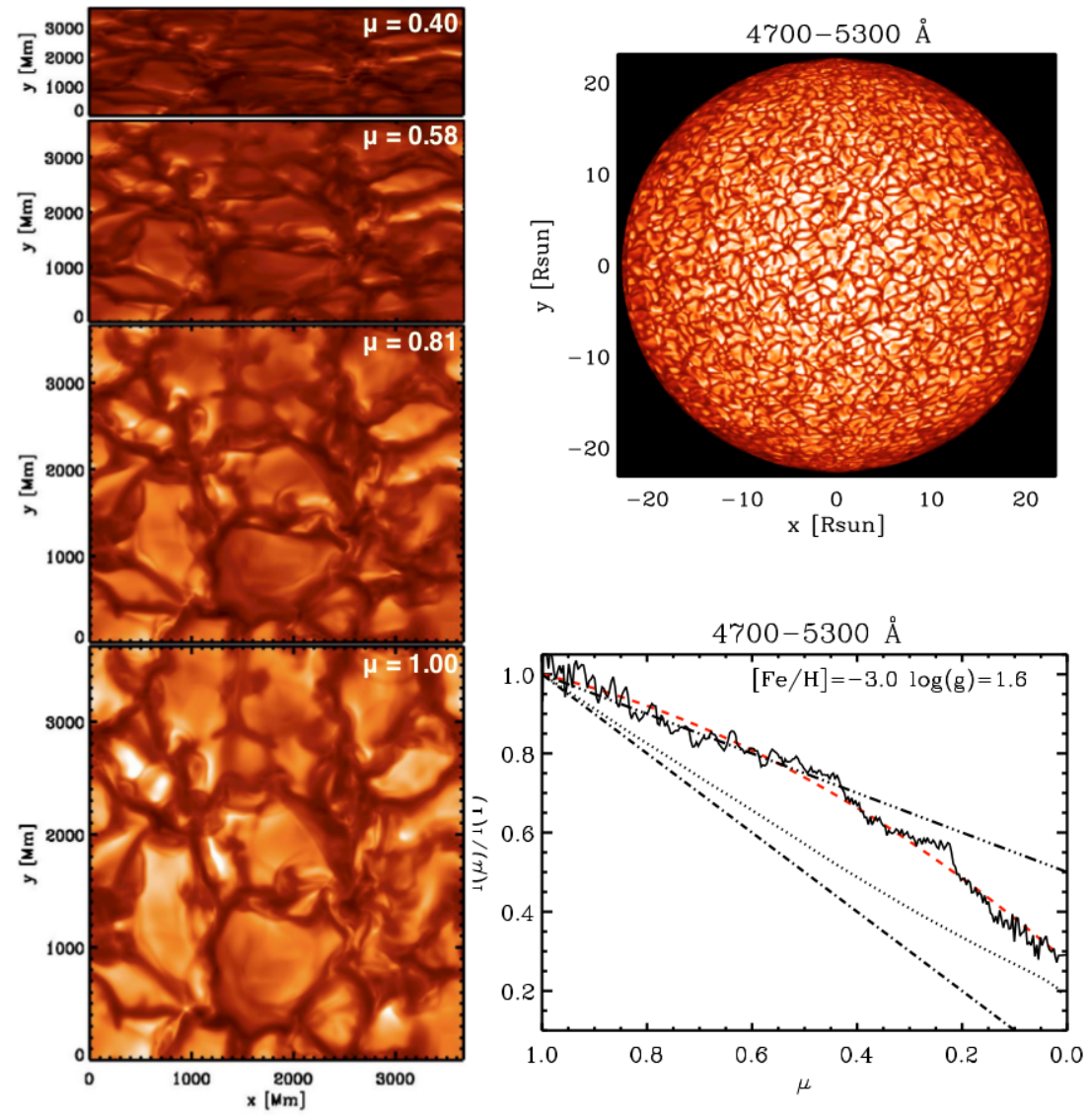}\\
 \end{tabular}
\caption[Semi-global model of RGB star]{\emph{Left:} Intensity maps at different inclined angles for a metal poor RGB star. \emph{Top right:} Synthetic stellar disk image obtained by tiling the intensity maps (left panel) ont a spherical surface \citep{2010A&A...524A..93C}. \emph{Bottom right:} CLV profiles from different models: RHD image (top right panel) azimuthal average (solid black line) with its numerical fit (red dashed line), a full limb darkened disk (dash-dotted line), and a partial limb darkened disk (triple dotted dashed line).}
              \label{fig:tiling}%
\end{figure}

The radius of a star is not a well-defined quantity since stars are gaseous spheres and do not have a well-defined edge \citep{2013ApJ...767....3D}. Optical and infrared interferometry has already proven to be a powerful tool for stellar astrophysics, in particular by providing fundamental stellar parameters such as Center to limb variations (CLVs) and radius as well as masses \citep[e.g. with CHARA interferometer][]{2012A&A...545A..17C}, which are compared to predictions by models of stellar evolution and stellar atmospheres.\\
Interferometric observables such as CLVs and stellar radii can also be obtained by the use of 3D RHD simulations. First, one needs to have an image of the stellar disk as a nearby observer would see it as required to extract the interferometric observables. However, the computational domain of each simulation only represents a small portion of the stellar surface (Fig.~\ref{fig:schema}). To overcome this limitation, and at the same time account for limb darkening effects (Fig.~\ref{fig:tiling}, left), we developed a technique \citep{2010A&A...524A..93C} that consists into computing intensity maps for different inclinations angles (with respect to the line of sight) and tile them onto a spherical surface (Fig.~\ref{fig:tiling}, top right). In addition to this, the statistical tile-to-tile fluctuations in the number of granules and in their shape is also taken into consideration. We used this method in several papers \citep{2010A&A...524A..93C, 2012A&A...540A...5C,2014A&A...567A.115C,2015A&A...576A..13C,2017A&A...597A..94C}.

\marginnote{\footnotesize \it Procyon and RGBs' stellar fundamental parameters}I report here two results concerning the RGB stars and the sub-giant star Procyon. 

\begin{itemize}
\item Red giant branch stars have evolved from the main sequence and are powered by hydrogen burning in a thin shell surrounding their helium core. Determine their fundamental parameters is very important because they are used as tracers of the morphology and evolution of the Galaxy in the framework of the GAIA mission and they are extensively used for spectroscopic elemental abundance analyses of distant stellar populations. In \cite{2010A&A...524A..93C}, we provided average limb darkening coefficients for different metallicities and wavelengths ranges (Fig.~\ref{fig:tiling}, bottom right). We found that the effect of convective-related surface structures depends on metallicity and surface gravity. Finally, we estimate 3D vs 1D corrections to stellar radii determination: RHD simulations are $\sim3.5\% $ smaller to $\sim1\%$ larger in the optical than 1D, and roughly 0.5 to 1.5$\%$ smaller in the IR. Even if these corrections are small, they are needed to properly set the zero point of effective temperature scale derived by interferometry and to strengthen the confidence of existing red giant catalogs of calibrating stars for interferometry. 
\item Procyon is one of the brightest stars in the sky and one of our nearest neighbours. It is therefore an ideal target for stellar astrophysics studies. The atmospheric parameters and the interferometric radius, which are used to define the stellar evolution model and the analysis of frequencies, depend strongly on the realism of the atmosphere and the exactness of the temperature gradient in the surface layers. In \cite{2012A&A...540A...5C}, we re-analyzed the interferometric and spectroscopic data at the different wavelengths to derive a new radius and provided limb-darkening coefficients in the optical as well as in the infrared. In addition to this, we computed also the asteroseismic radius. Eventually, we provided also the resulting effective temperature and surface gravity. 
\end{itemize}

\subsection{Granulation and interferometry}

\begin{figure}[!h]
   \centering
  \begin{tabular}{cc}
  \includegraphics[width=1.0\hsize]{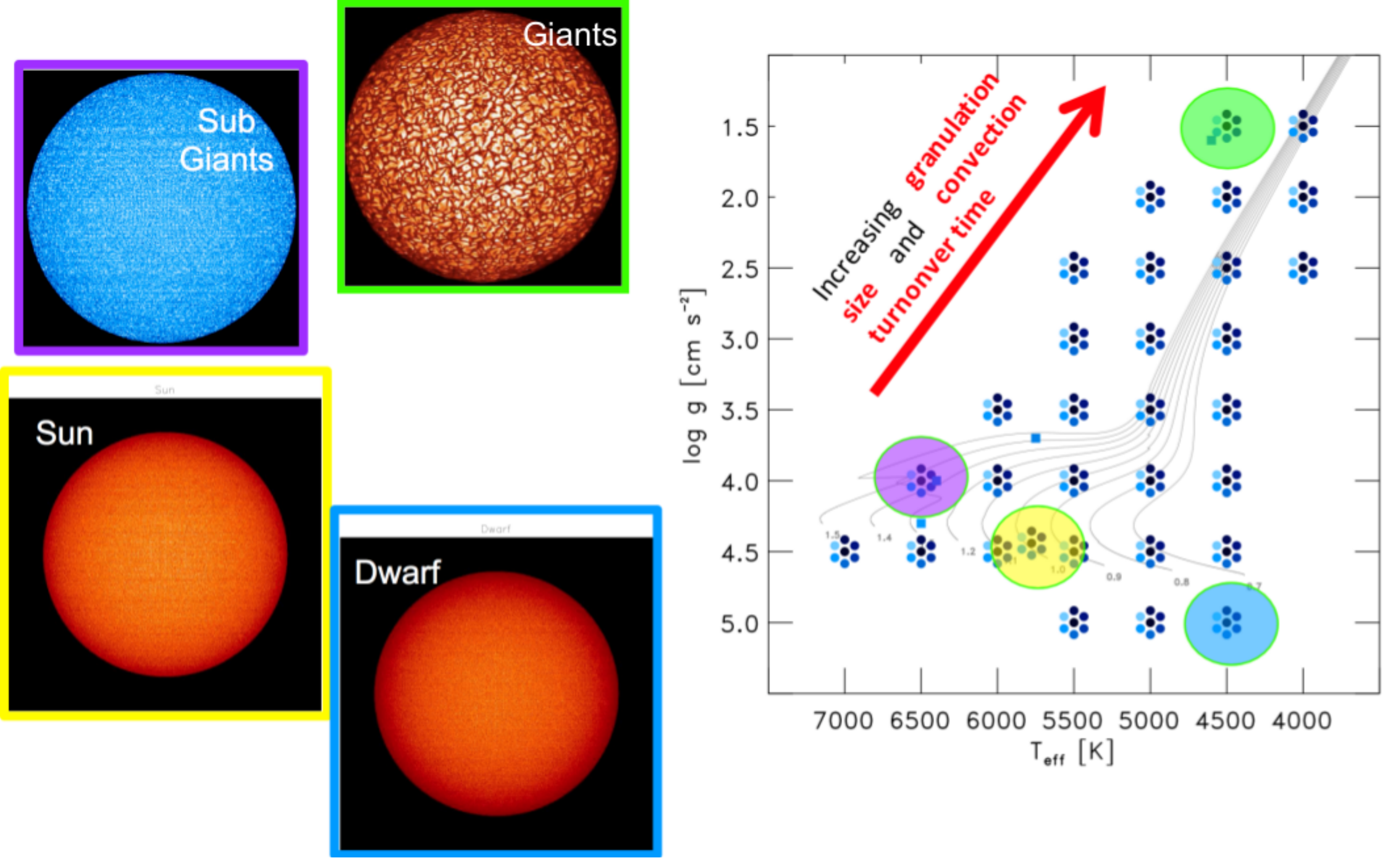}\\
 \end{tabular}
\caption[Semi-global model across the H-R diagram]{\emph{Left:} Synthetic images of the stellar disk for some 3D RHD simulations \citep{2017A&A...597A..94C,2014A&A...567A.115C,2012A&A...540A...5C,2010A&A...524A..93C} associated with their positions in Hertzsprung-Russell diagram (\emph{right}).}
              \label{fig:3Dgran}%
\end{figure}

The size of the convective cells is correlated to the pressure scale height at optical-depth unity \citep{2001ASPC..223..785F}, confirmed later by \cite{2013A&A...557A...7T}. The pressure scale height is
defined as

\begin{eqnarray}
\mathcal{H}_{\mathrm{p}}= \frac{k_B T_{\mathrm{eff}}}{mg},
\end{eqnarray}

where $g$ is the surface gravity, $k_B$ is the Boltzmann constant and
$m$ is the mean molecular mass ($m=1.31\times m_{\rm{H}} = 1.31 \times
1.67 \times10^{-24}$~grams, for temperatures lower than 10000~K). In
the above expression, $\mathcal{H}_{\mathrm{p}}$ has the dimension of
length. Stars with low surface gravity have more diluted atmosphere
and lower surface temperature while more compact objects are
hotter. Fig.~\ref{fig:3Dgran} shows the synthetic images of stars with
different stellar parameters and thus different granulation pattern. Connected to this, also the turnover timescales are also related to the granulation pattern as displayed in the Figure.

\marginnote{\footnotesize \it Interferometric closure phases}The role of long-baseline interferometric observations is to investigate the dynamics of granulation as a function of stellar parameters: thanks to the higher angular resolution, interferometry is the ideal tool for exploring stellar convection in term of cellÕs size, intensity contrast and temporal variations. In \cite{2014A&A...567A.115C}, we characterised the granulation on different type of stars using interferometric observables. For this purpose, we used the interferometric observable called "closure phase", which is defined as the phase of the triple product
(or bispectrum) of the complex visibilities on three baselines that form a closed loop joining (at least) three stations. This procedure removes the atmospheric contribution, leaving the phase
information of the object morphology unaltered \citep{2007NewAR..51..604M}. To sum up, values of closure phases different from zero$^{\circ}$ or $\pm180^{\circ}$ means that the observed object is not centrosymmetric. \\
At the end, the characteristic size distribution on the stellar surface can be derived from the closure phase: the contribution of small-scale convection-related surface structures increases with frequency \citep[see Section~\ref{sizessect}, ][]{2010A&A...515A..12C}. \cite{2014A&A...567A.115C} reported that the granulation has a very clear signature in all type stars and for different interferometric instruments depending on the wavelength range sampled and on the UV-coverage used. In particular, MIRC instrument mounted at CHARA interferometer is the most appropriate instrument because it combines good UV-plane coverage and long baselines (ie, high spatial resolving power together with very good coverage of the stellar surface). The signature of the convective-related surface structures stars to be relevant from spatial frequencies corresponding to the third lobe (i.e., structures up to about 1/10 of the stellar radius). \\
The first confirmation of our predictions came with MIRC observations of three red giant branch (RGB) stars: HD 197989, HD 189276, and HD 161096. All of them with apparent radius between 2 and 4.5 mas. We detected departures from the centrosymmetric case for all three stars with the tendency of a greater effect for the stars with lower surface gravity, even though our sample is limited. This qualitatively means that the more the star evolves, the more significant the size of the granules becomes with respect to the disk size. This idea is supported by previous work showing even larger departures from centrosymmetry for very evolved stars such as AGBs \citep{2006ApJ...652..650R, 2010A&A...511A..51C, 2016A&A...587A..12W} and RSGs \citep{2010A&A...515A..12C,2018A&A...614A..12M}.

\subsection{Exoplanet transits}

Among the different methods used to detect exoplanets, the transit method is a very successful technique: 2487 over 3021 transit candidates have been confirmed with this technique \citep[as of November 2018 from http://exoplanets.org, ][]{2011PASP..123..412W}. A transit event occurs when the planet crosses the line of sight between the star and the observer, thus occulting part of the star. This creates a periodic dip in the brightness of the star. The typical stellar light blocked is $\sim1\%$, 0.1$\%$, and 0.01$\%$ for Jupiter-, Neptune- and Earth-like planets transiting in front of a Sun-like star, respectively \citep{1984Icar...58..121B}, making the detection very challenging, in particular for Earth-like planets. During the transit, the flux decrease is proportional to the squared ratio of planet and stellar radii. For sufficiently bright stars, the mass can also be measured from the host star's radial velocity semi-amplitude \citep{2012A&A...538A...4M}. When
the mass and radius of an exoplanet are known, its mean density can also be deduced and provide useful information for the physical formation processes. Today and in the near future, the prospects for planet detection and characterization with the transiting methods are excellent with access to a large amount of data coming, for instance, from the NASA missions Kepler \citep{2010Sci...327..977B}, TESS \citep[Transit Exoplanet Survey Satellite,][]{2010AAS...21545006R}, and JWST\footnote{https://www.jwst.nasa.gov/} or from the ESA missions PLATO 2.0 \citep[PLAnetary Transits and Oscillation of stars,][]{2014ExA...tmp...41R} and CHEOPS \citep[CHaracterizing ExOPlanet Satellite, ][]{2013EPJWC..4703005B}. \\

\begin{figure}[!h]
\centering
\begin{tabular}{ccc}
\includegraphics[angle=0,width=0.95\hsize]{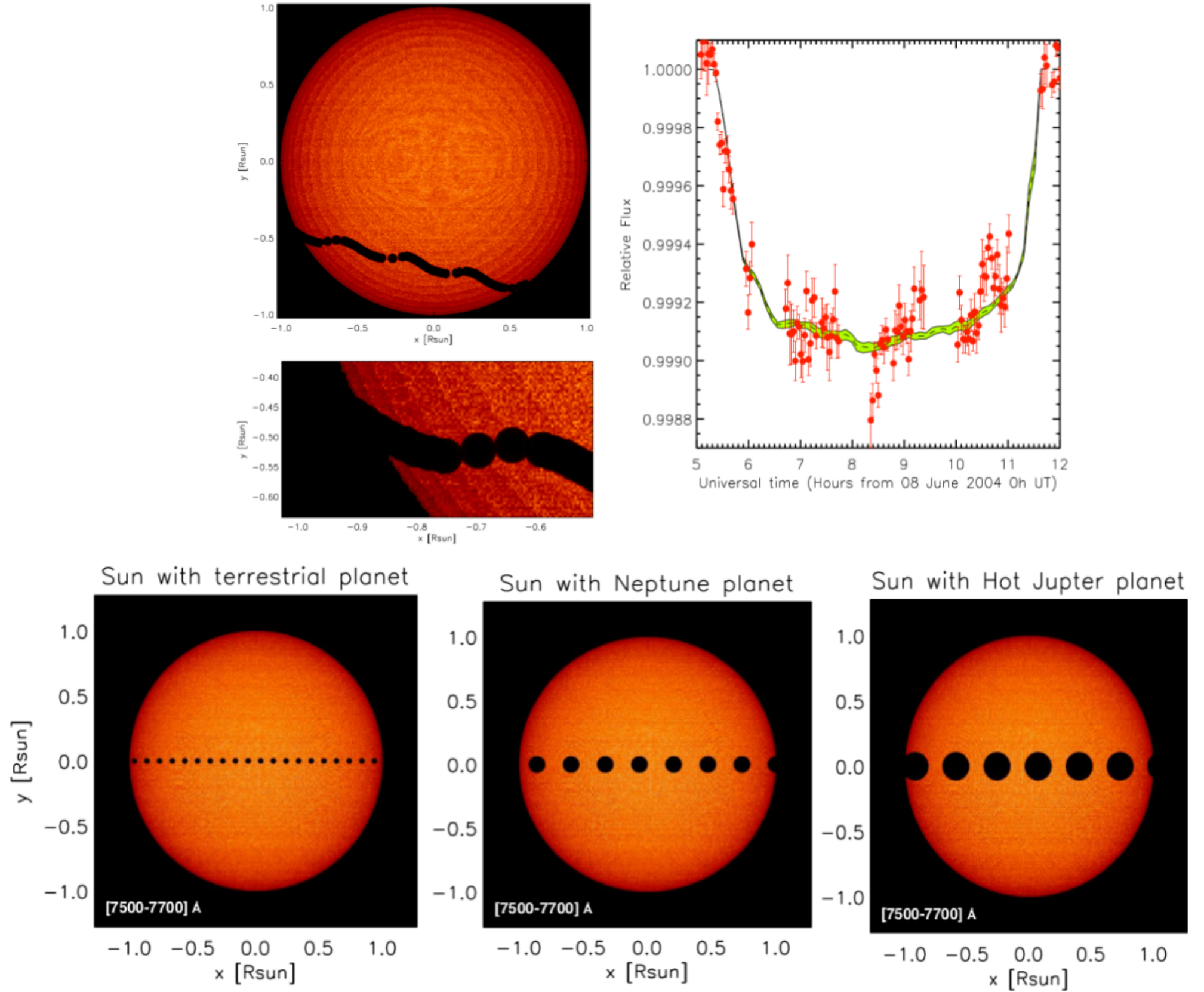}
\end{tabular}
\caption[Synthetic transits]{\emph{Top left panel}: Synthetic image (with enlargement) of the solar disk in the visible and the different positions of the transit of Venus in 2004 as seen from the ACRIMSAT satellite. Venus' unusual trajectory is induced by the satellite's orbit. \emph{Top right panel}: Light curve of the Venus transit, photometric observations are reported in red with error bars \citep{2006ApJ...641..565S}, the dotted black line is the best matching curve from 3D simualtions, and the green color indicates the highest and lowest values due to the changes in granulation in the synthetic Sun \citep{2015A&A...576A..13C}. \emph{Bottom panels:} Synthetic transits in front of the Sun for several types of planets to study the impact of granulation on the measurement of the planetary radius \citep{2017A&A...597A..94C}.}
\label{transitfig}
\end{figure}

\marginnote{\footnotesize \it Granulation noise in planet transits}However, with improved photometric precision, additional sources of noise that are due to the presence of stellar surface inhomogeneities such as granulation, will become relevant, and the overall photometric noise will be less and less dominated by pure photon shot noise. I carried out an in-depth study of a very particular transit, that of Venus in 2004. To do this, I used the spherical tile imaging method \citep[see Section~\ref{sec:param}, ][]{2010A&A...524A..93C} to construct the semi-global model of the Sun and then we modelled the light curve of the Venus transit, seen from the ACRIMSAT satellite (Fig.~\ref{transitfig}, top left panel), using several realizations of the solar disk (to represent the temporal variation of the granulation). Then, we compared the synthetic curves with the observations and found a very good agreement, both in terms of depth and entry/exit slopes of the transit (top right panel). In the end, the granulation pattern causes fluctuations in the transit light curve that can cause intrinsic uncertainty (due to stellar variability) on accurate measurements of transits \citep{2015A&A...576A..13C}. \\
 	Following the method described for the transit of Venus, I evaluated the impact of granulation at different wavelengths (from optical to IR) for several planet/star systems: I simulated the transits of three prototype planets a hot Jupiter, a hot Neptune, and a terrestrial planet, orbiting around a solar and K-dwarf type stars. For the first time from the point of view of the star 	(Fig.~\ref{transitfig}, bottom panels). We  demonstrated that granulation has a significant effect on the depth of the light curve during the transit, and consequently on the determination of the planetary radius (up to 0.90$\%$ and $\sim$0.5$\%$ for terrestrial and gaseous planets, respectively). We also showed that larger (or smaller) orbital inclination angles with respect to values corresponding to transit at the stellar center display a shallower transit depth and longer ingress and egress times, but also granulation fluctuations that are correlated to the center-to-limb variation: they increase (or decrease) the value of the inclination, which amplifies the fluctuations. 
	
	In conclusion, the granulation has to be considered as an intrinsic uncertainty (as a result of stellar variability) on the precise measurements of exoplanet transits of planets. The full characterization of the granulation is essential for determining the degree of uncertainty on the planet parameters. In this context, the use of 3D RHD simulations is important to measure the convection-related fluctuations. This can be achieved by performing precise and continuous observations of stellar photometry and radial velocity, as we explained with RHD simulations, before, after, and during the transit periods \citep{2017A&A...597A..94C}.

\textit{\textcolor{blue}{The paper, "Measuring stellar granulation during planet transits" by Chiavassa et al. 2017, is attached at the end of the chapter.}}.

\clearemptydoublepage

\section{Evolved cool stars}

\lettrine[lines=3,nindent=4pt]{W}{e now} move to the results concerning the evolved stellar branch of the H-R diagram.

\subsection{Let's compute a RSG simulation}

In {\sc CO5BOLD} (see Section~\ref{hydro}), the most important parameters \citep{2011A&A...535A..22C} that determine the type of the simulated star are:
\begin{itemize}
\item the input luminosity into the core
\item the stellar mass that enters in the equation for the gravitational potential
\item the abundances that are used to create the tables for the equation-of-state and the opacities.
\end{itemize}

\begin{figure}[!h]
\centering
\begin{tabular}{ccc}
\includegraphics[angle=0,width=0.95\hsize]{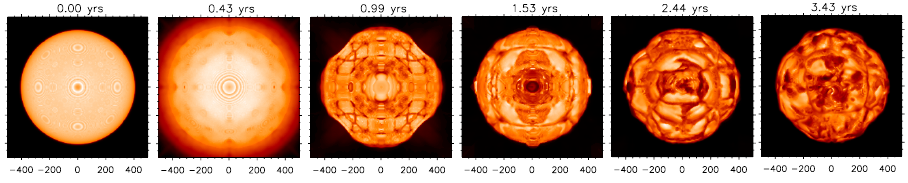}
\end{tabular}
\caption[Computing 3D RHD simulations of RSG stars]{Gray intensity on one side of the computational cube from the initial sequence of a 3D RHD simulation of an RSG. The axes are in solar radii. The artifacts caused by the mismatch between the spherical object and the Cartesian grid become less evident with time passing \citep{2011A&A...535A..22C}.}
\label{fig:3dcomputation}
\end{figure}

The initial model is produced starting from a sphere in hydrostatic equilibrium with a weak velocity field inherited from a previous model with different stellar parameters (Fig.~\ref{fig:3dcomputation}). After some time, the limb-darkened surface without any convective signature appears but with some regular patterns due to the numerical grid. The central spot, quite evident at the beginning of the simulation, vanishes completely when convection becomes strong. After several years of stellar time, a regular pattern of small-scale convection cells develops and, after cells merge the average structures, it becomes big and the regularity (due to the Cartesian grid) is lost. The intensity contrast grows with time.

\marginnote{\footnotesize \it Radiation transport in RHD simulations}The radiation transport for the simulations of evolved stars employs a short-characteristics
method, and, to account for the short radiative time scale, several
(typically 6 to 9) radiative sub$-$steps are performed per
global step. The simulations can be computed: (i) either using a gray frequency dependance of the radiation field, which ignores the frequency dependence, based on Rosseland mean opacities calculated merging high-temperature OPAL \citep{1992ApJ...397..717I} data and low-temperature
PHOENIX \citep{1997ApJ...483..390H} at around 12\ 000K; (ii) or using a multi-group scheme \citep{1994A&A...284..105L,2004A&A...421..741V}, where the frequencies that reach monochromatic optical depth unity within a certain depth range of the model atmosphere will be put into one frequency group. The RHD simulations employing the latter method have typically five wavelengths groups sorted according to the run of the monochromatic optical depth in a corresponding MARCS \citep{2008A&A...486..951G} 1D model with a smooth transition to the Rosseland mean (OPAL opacities) in the optically thick regime.

Once the RHD simulation is relaxed, the snapshots are used for detailed post-processing treatment to extract interferometric, spectrophotometric, astrometric, and imaging observables using {{\sc Optim3D}}.

\subsection{Simulations characteristics and timescales}

RHD simulations of evolved stars show a very heterogeneous surface caused by the dynamical granulation. The emerging intensity is related to layers where waves and shocks dominate together with the variation in opacity through the atmosphere. Small-amplitude acoustic waves are produced in the convective envelope by non-stationary convective flows with significant Mach numbers (e.g., 5 or even larger). These waves can travel outward in the convective envelope and even into the convectively stable atmosphere where they are compressed and amplified (due to the lower temperature and sound speed) and further amplified (due to the lower density). Here, they turn into shocks giving rise to a dynamical pressure larger than the gas pressure \citep{2011A&A...535A..22C,2017A&A...600A.137F}.

	RHD simulations pulsate by themselves and do not have any dynamic boundary condition, the hydrodynamical equations include the advection of momentum, which, after averaging over space and time, gives the dynamical pressure \citep{2017A&A...600A.137F}. AGB models show more extend and varying structures in the near-surface than RSG (i.e., this is a consequence of the higher mass), and, convective velocities in RSGs are too low to reach escape velocity and contribute to the mass-loss mechanism. RSG and AGB simulations are both characterized by large convective cells and strong shocks, however, AGBs have in general even larger scales with shocks pushing the mass much further out.

\begin{figure}[!h]
\centering
\begin{tabular}{c}
\includegraphics[angle=0,width=0.6\hsize]{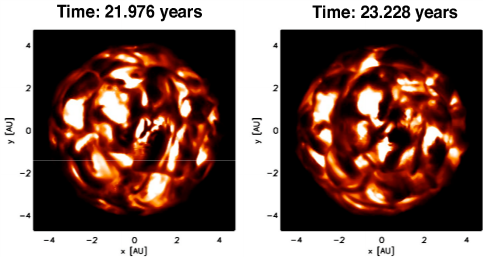}\\
\includegraphics[angle=0,width=0.68\hsize]{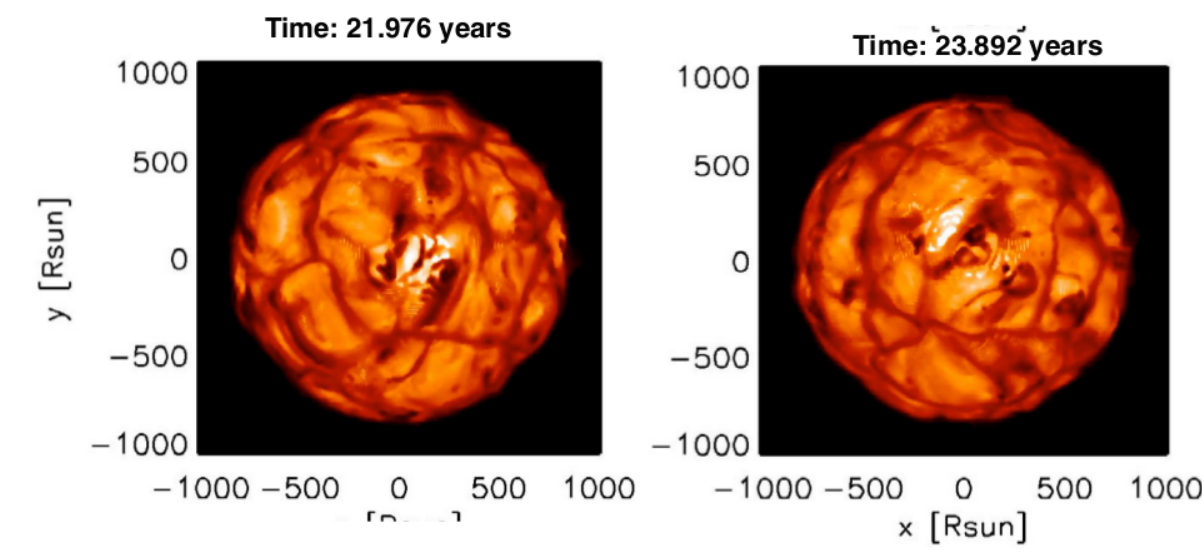}\\
\includegraphics[angle=0,width=0.6\hsize]{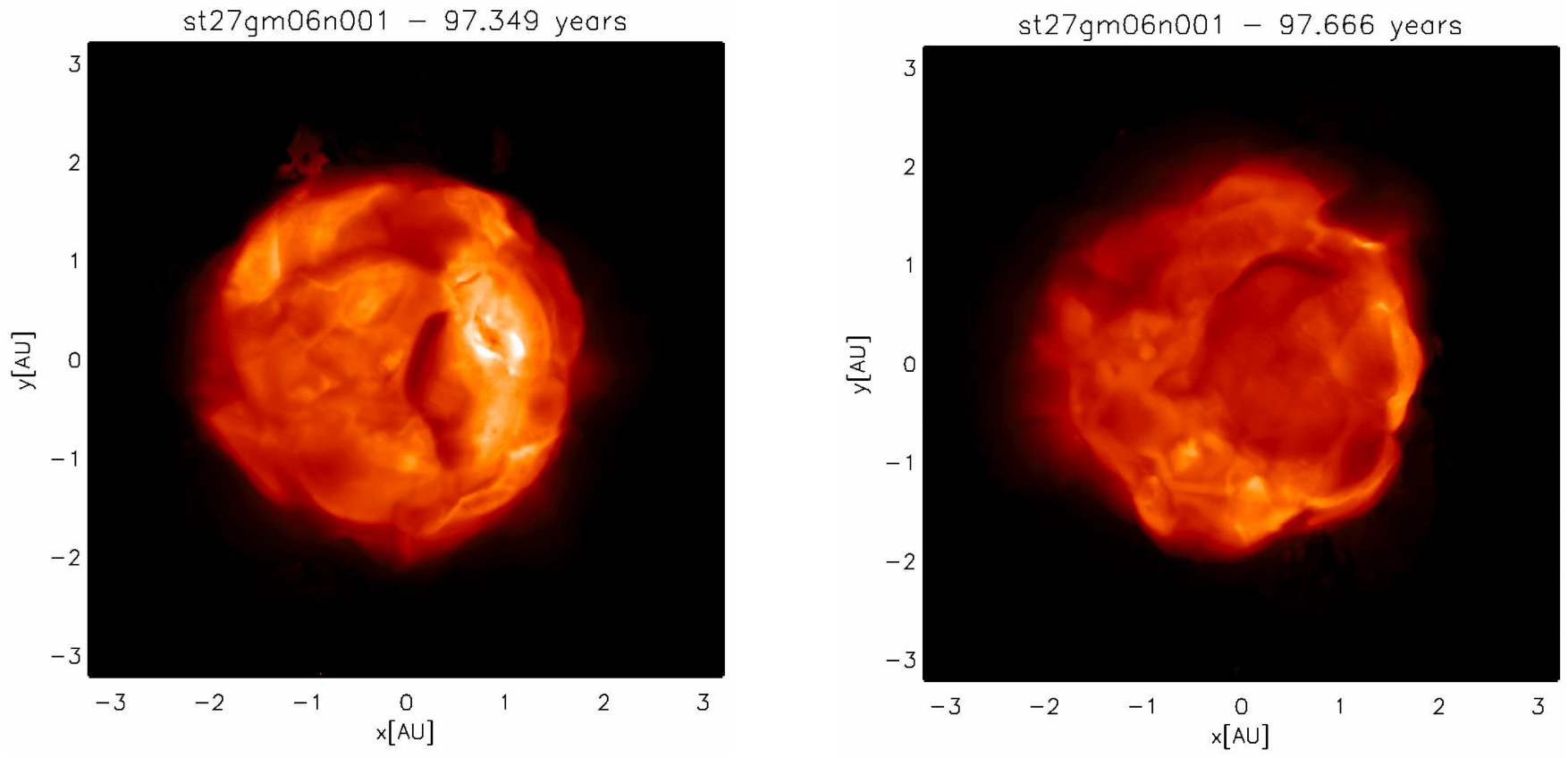}
\end{tabular}
\caption[Timescales in evolved stars]{\emph{Top panels:} Maps of the linear intensity of an RSG simulation in the Gaia $G$ band \citep[][]{2018arXiv180409368E}. Each panel corresponds to a different snapshot whose time is indicated \citep{2011A&A...528A.120C}. \emph{Central panels:} Maps of the linear intensity of the same simulation above but in the H band \citep[IONIC filter, $\sim$1.6 $\mu$m, ][]{2009A&A...506.1351C}. \emph{Bottom panels:} Squared root intensity maps of an AGB simulation in the Gaia $G$ band \citep{2018arXiv180802548C}.}
\label{fig:timescales}
\end{figure}

\begin{table}[!ht]

\caption{Typical stellar parameters for RHD simulations of RSG \citep{2011A&A...535A..22C} and AGB \citep{2017A&A...600A.137F} stars.}
\label{simus}  
\smallskip
\begin{center}
{\footnotesize
\begin{tabular}{cccccccccc}
\tableline
\noalign{\smallskip}

Type & Numerical & $M_{\mathrm{pot}}$  & $L$ &  $T_{\rm{eff}}$    \\
                  & resolution & $\!\!\!$[$M_\odot$]$\!$  & $\!\!\!$[$L_\odot$]$\!$ & $\!\!\!$$[\rm{K}]$$\!$ \\

\noalign{\smallskip}
\tableline
\noalign{\smallskip}
RSG & up to $401^3$ &  5--12   &  $\approx$30000--90000 & $\approx$3300--4000 \\
AGB & up to $401^3$ & 1 & $\approx$5000--10000 & $\approx$2500--2900 \\
\tableline
\noalign{\smallskip}
\noalign{\smallskip}
\noalign{\smallskip}
Type & $R_{\star}$ &  $\log g$ & Rotation & Longest & Detected   \\
              & $\!\!\!$[$R_\odot$]$\!$ &  $\!\!\!$[cgs]$\!$  & simulations? &simulation [y] & Pulsation? \\

\noalign{\smallskip}
\tableline
\noalign{\smallskip}

RSG &  $\approx$400-900 & $\approx$$-$0.45 -- 0.00  & yes & 30 & not yet\\
AGB &  $\approx$300--550 & $\approx$$-$1.00 -- $-$0.50 & yes & 30 & yes \\
\noalign{\smallskip}
\tableline
\end{tabular}
}
\end{center}
\end{table}

\marginnote{\footnotesize \it Two main temporal scales depending on wavelength probed}The temporal timescales of the granulation pattern are different with respect to the spectral range probed. An example is reported in Fig.~\ref{fig:timescales}. The wavelength dependence is striking, but it is also important to note that going from the infrared to the optical, there is a relevant increase of the intensity contrast\footnote{In average, the brightest areas exhibit an intensity $\sim$50 times or larger than the dark ones in the optical and up to $\sim$10 times in the infrared. Probing very narrow wavelength filters close to particular spectral line centers may increase these values.} as well as the number and complexity of surface structures. Concerning the timescales, the simulations are characterized by two principal characteristic time scales linked directly to the stellar dynamical effects:

\begin{itemize}
\item the surface is covered by a few large convective cells with a size of $\approx60\%$ of the stellar radius (top row of Fig.~\ref{fig:timescales}) that evolve on a time scale of years \citep{2009A&A...506.1351C}. This is visible in the infrared, and particularly in the H band where the H$^-$ continuous opacity minimum occurs and consequently the continuum-forming region is more evident;
\item in the optical region \citep[central and bottom row of the Figure, and ][]{2011A&A...528A.120C,2017A&A...600A.137F,2018arXiv180802548C}, short-lived (a few weeks to a few months) small-scale structures appear. They result from the opacity contribution and dynamics at optical depths smaller than 1 (i.e., further up in the atmosphere with respect to the continuum-forming region), as well as from the higher sensitivity of the blackbody radiation to the temperature inhomogeneities. It must be noted that also the numerical resolution of the simulation plays a role for the size and number of these small structures.
\end{itemize}

\subsection{Spatially unresolved surfaces: measuring convection cycles with velocity fields at high spectral resolution}

\begin{figure}[!h]
\centering
\begin{tabular}{c}
\includegraphics[angle=0,width=0.5\hsize]{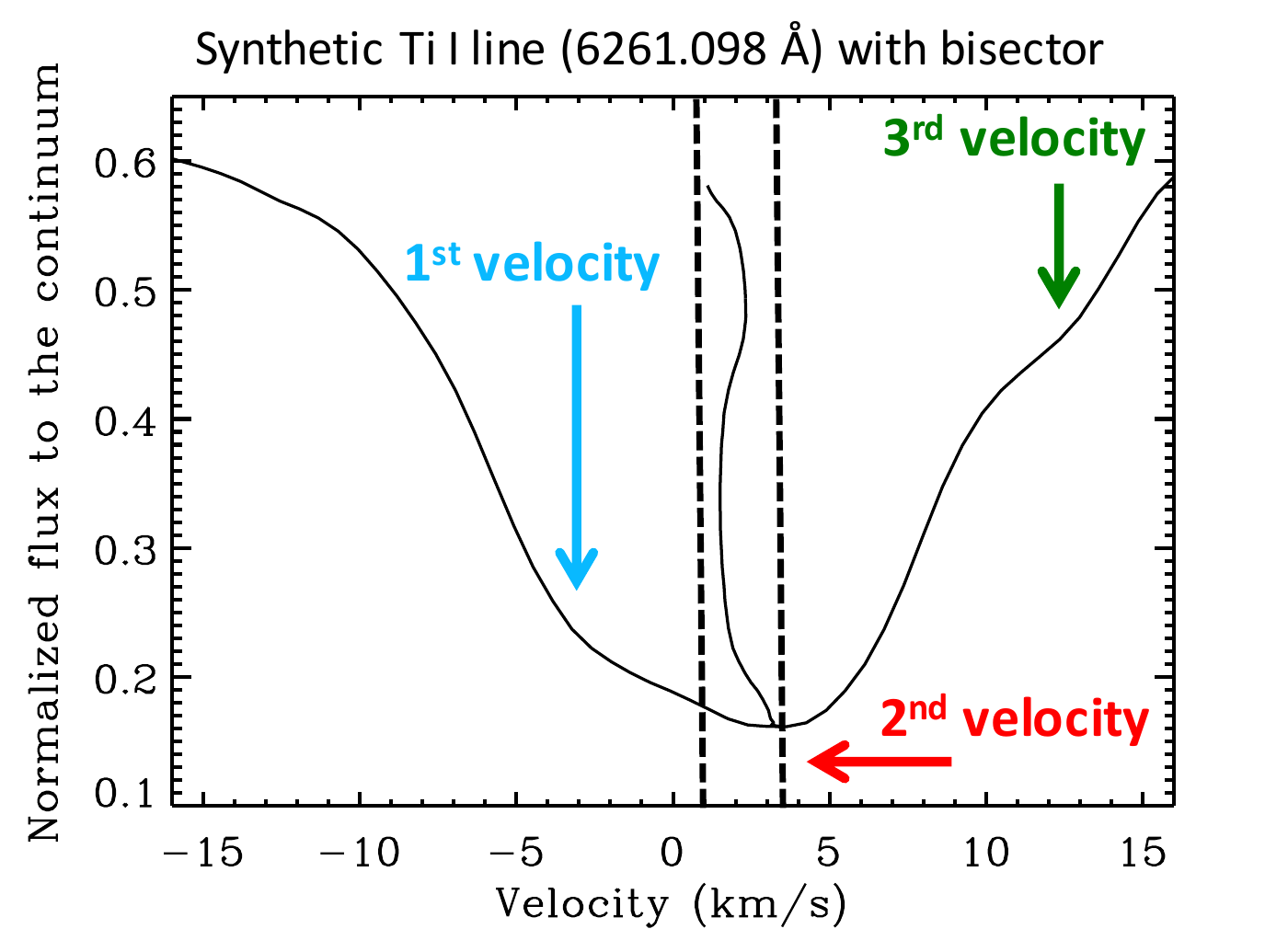}
\includegraphics[angle=0,width=0.5\hsize]{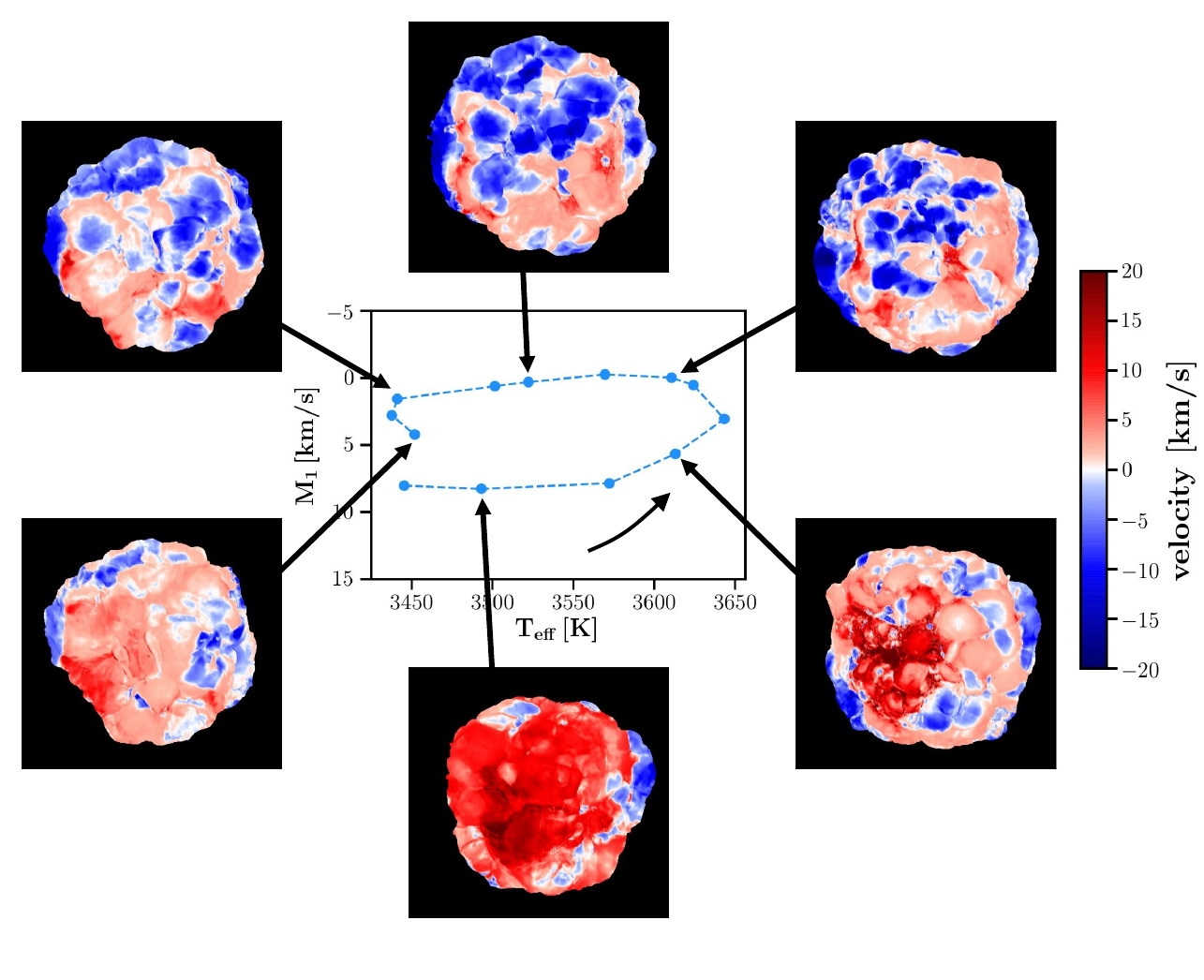}
\end{tabular}
\caption[Velocity fields in evolved stars]{\emph{Left panel:} Synthetic spectrum of the Ti I line at 6261.11 \AA\ for one snapshot of a 3D RHD simulation of an RSG star \citep{2015ASPC..497...11C}. The vertical dashed line shows the spanned velocities of the line bisector. The different arrow and colors displays the positions of different velocity components which contribute to the shape of the line. \emph{Right panel:} Velocity maps for different snapshots of a RSG simulation during a convection cycle (central part of the panel). The velocity is weighted with the contribution function and red/blue colors correspond to falling/approaching material, respectively (Kravchenko, Chiavassa, Van Eck et al., to be submitted).}
\label{fig:vel}
\end{figure}

	RHD simulations provide a self-consistent ab-initio description of the non-thermal velocity field generated by convection, shock waves, and overshoot that manifests itself in spectral line shifts and changes in the equivalent width (Fig.~\ref{fig:vel}, left panel). The shape of the optical Ti I line at 6261.11 \AA\ , taken as an example here, constitutes of more than one velocity component that contributes through the different atmospheric layers where the line forms. As a consequence, the line bisector\footnote{It is the locus of the midpoints of the line. A symmetric profile has a straight vertical bisector, while the "C"-shaped line bisector reveals asymmetries.} is not straight and span values up to 5 km/s on a temporal scale of few weeks \citep[as qualitatively seen in the prototypical RSG star $\alpha$~Ori by ][]{2008AJ....135.1450G}.
As the vigorous convection is prominent in the emerging flux, the radial velocity measurements for evolved stars are very complex and need a sufficiently high spectral resolution to possibly disentangle all the sources of macro-turbulence.  \\
In this context, Kateryna Kravchenko (the PhD student I co-supervise with S. Van Eck) is working on the tomographic method in the framework of evolved stars. The method allows to recover the distribution of the component of the velocity field projected on the line of sight at different optical depths in the stellar atmosphere \citep{2000A&A...362..655A,2001A&A...379..305A,2001A&A...379..288A}.  \cite{2018A&A...610A..29K} introduces the recently updates to the method as well as the implementation in {{\sc Optim3D}} for the calculation of the contribution function in 3D RHD simulation of RSG stars. Kateryna successfully managed to show that in 3D simulations, the spectral lines do not form in a restricted range of reference optical depths as in 1D model atmospheres, but they spread over different optical depths due to the non-radial convective mouvements characterising the stellar atmosphere. In addition to this, she managed also to recover RHD simulation velocity field dependance across the atmosphere with this method. \\
	The latter opens a new doorway for the study of stellar dynamical cycles in evolved stars, and in particular RSGs. A first example is reported in Kravchenko et al. 2018 (to be submitted) where the tomographic method help to interpret the long-term monitoring (almost 7 years with HERMES spectrograph) photometric variability of the RSG $\mu$ Cep. The tomographic method denoted, in the observations, the characteristic of the convective \marginnote{\footnotesize \it Convective turn-over time in observations} turn-over of the material in the stellar atmosphere \citep[also knows as hysteresis loop][Fig.~\ref{fig:vel}, right panel]{2008AJ....135.1450G} and RHD simulations qualitatively explain this behaviour: the velocity maps in (same Figure) reveal upward and downward motions of matter extending over large portions of the stellar surface. The relative fraction of upward and downward motions is what distinguishes the upper from the lower part of the hysteresis loop, its top part (zero velocity) being characterized by equal surfaces of rising and falling material. The bottom part of the hysteresis loop occurs, as expected, when the stellar surface is covered mostly by downfalling material.

\subsection{Spatially resolved surfaces: probing stellar parameters and surface details}\label{sizessect}

\begin{figure}[!h]
\centering
\begin{tabular}{ccc}
\includegraphics[angle=0,width=0.95\hsize]{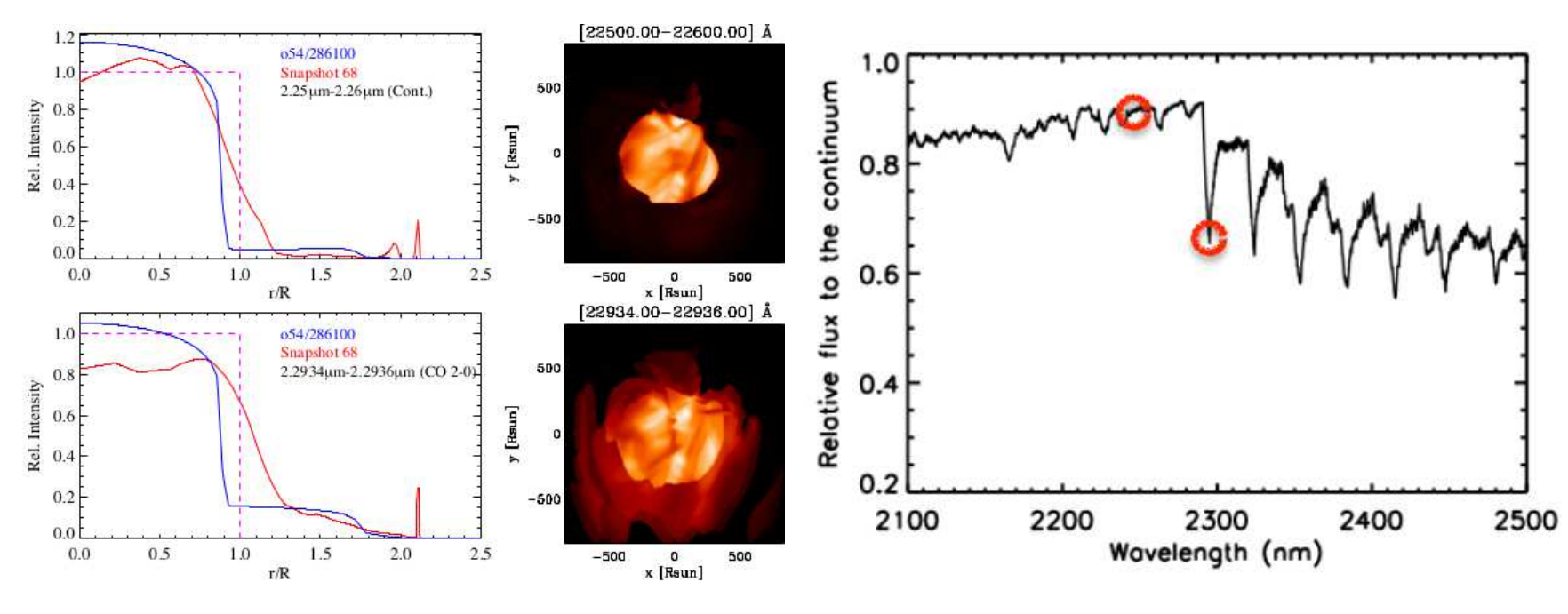}
\end{tabular}
\caption[Atmospheric extension with interferometry]{\emph{Left panel:} Intensity profiles (in red) and intensity maps of a 3D RHD simulation of an AGB star in two spectral intervals in the K band \citep{2016A&A...587A..12W}. \emph{Right panel:} flux calculated for the same simulation. The red circles represent approximately the position of the selected intervals in the K-band.}
\label{fig:agbinterfero}
\end{figure}

\begin{figure}[!h]
\centering
\begin{tabular}{ccc}
\includegraphics[angle=0,width=0.95\hsize]{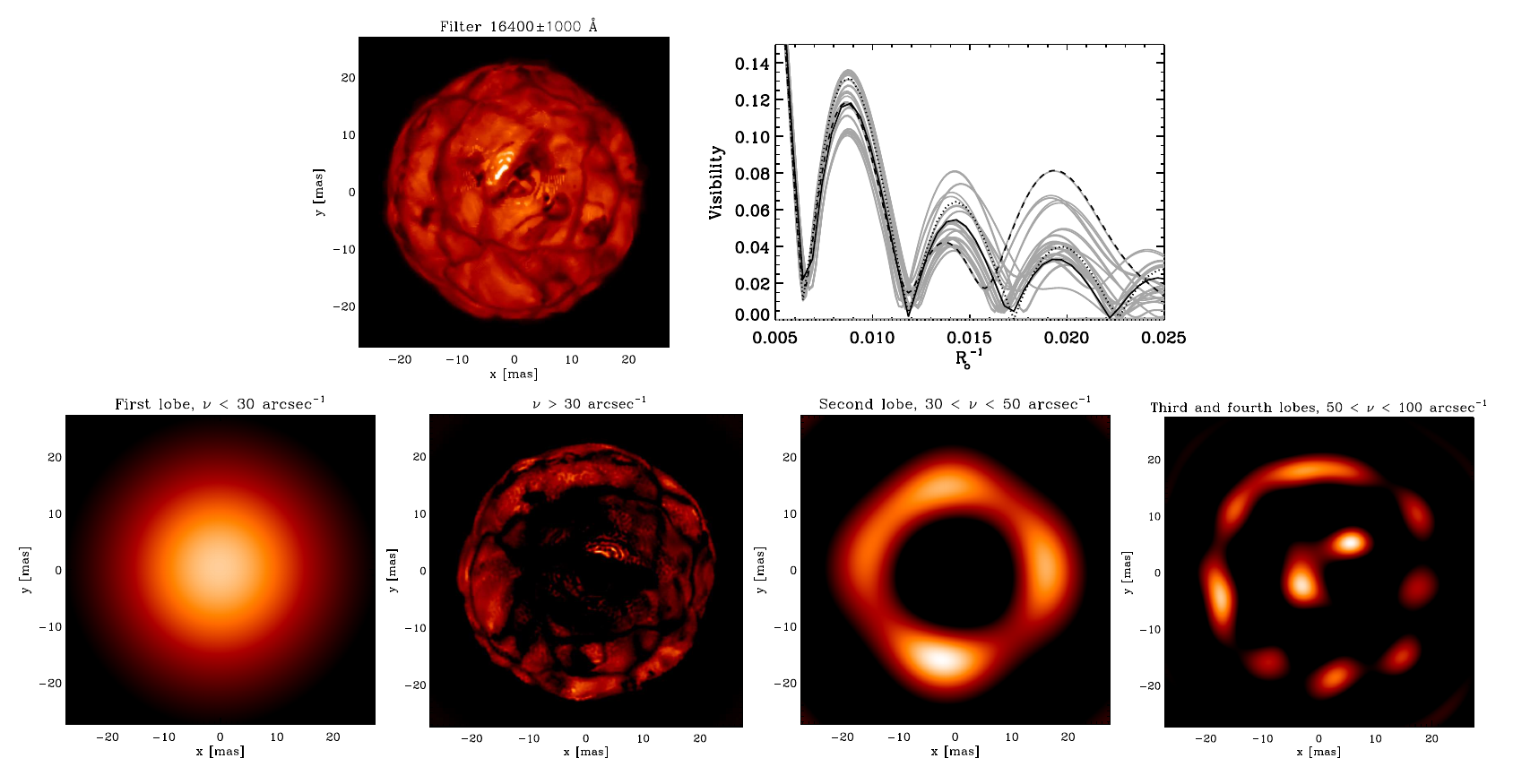}
\end{tabular}
\caption[Characteristic size of convective cells]{Characteristic size of convective cells on the prototypical RSG Betelgeuse \citep{2009A&A...506.1351C,2010A&A...515A..12C}. \emph{Top left panel:} Synthetic intensity map in the H-band \citep{2009A&A...506.1351C}. \emph{Top right panel:} visibility curves (grey and black lines) computed from the image on the left for different rotation angles (36 in total with a step of 5 degrees). The dashed curve is the LD visibility while the black curve is the visibility for an uniform disc. \emph{Low panels:} Filtered intensity maps at different spatial frequencies corresponding to the different lobes of the visibility curve \citep{2010A&A...515A..12C}.}
\label{fig:sizes}
\end{figure}

Interferometry importance for evolved stars is two-folds: (i) it allows the determination of their stellar parameters, and (ii) affords the direct detection and characterization of the convective pattern related to the surface dynamics. 

\begin{itemize}
\item Point (i). The direct measurement of stellar angular diameters has been the principal goal of most attempts with astronomical interferometers since the pioneering work \citep{1921ApJ....53..249M}. Nowadays with the advent of Gaia, for stars of known distance the angular diameter becomes of paramount importance to yield the stellar radius and eventually to the absolute magnitude. These quantities are essential links between the observed properties of stars and the results of theoretical calculations on stellar structure and evolution. Recent pioneering survey works \citep{2012A&A...540L..12W,2013MNRAS.434..437C,2014A&A...566A..88A,2015A&A...575A..50A,2017A&A...597A...9W}, in which I took part, characterized the fundamental parameters and atmospheric extensions of evolved stars in our neighbourhood using AMBER instrument (now decommissioned) at VLTI. In particular, the last two papers observed a linear correlation between the visibility ratios of observed RSGs and the luminosity and surface gravity, indicating an increasing atmospheric extension with increasing luminosity and decreasing surface gravity, indirectly supporting a mass-loss scenario of a radiatively driven extension caused by radiation pressure on Doppler-shifted molecular lines. These results are confirmed for AGB stars \citep[Fig.~\ref{fig:agbinterfero},][]{2016A&A...587A..12W} where the atmospheric extension are detected and explained by the RHD simulations for a sample of interferometric observations. The latter support the mass-loss scenario of pulsation- and shock-induced dynamics that can levitate the molecular atmospheres of Mira/AGB variables to extensions that are consistent with observations.
\item Point (ii). Two main observables are used in interferometry: the visibility and the closure phases. The combination of both, plus a good enough coverage of the Fourier plane, contributes to the image reconstruction of the observed targets. Visibilities measure the surface contrast of the source and are primarily used to determine the fundamental stellar parameters and the limb-darkening. Closure phases \marginnote{\footnotesize \it Observing the details of the large granule means to understand the mass-loss}combine the phase information from three (or more) telescopes and provide direct information on the morphology of the source. For a correct interpretation of the observations, it is necessary to simultaneously explain both observables with the same model. \cite{2009A&A...506.1351C,2010A&A...515A..12C} detected and measured the characteristic sizes of convective cells on the RSG star $\alpha$~Ori (Fig.~\ref{fig:sizes}) using visibility measurements in the infrared and managing to explain the observation of $\alpha$~Ori from the optical to the infrared region using RHD simulations. They showed that its surface is covered by a granulation pattern that, in the H and K bands, shows structures with small to medium scale granules (5-15 mas, while the size of the star at these wavelengths is $\sim$44 mas) and a large convective cell ($\sim$30 mas). Another result concerns the first reconstructed images with AMBER/VLTI of the massive evolved star VX~Sgr \citep{2010A&A...511A..51C}. The authors used RHD simulations of RSG and AGB to probe the presence of large convective cells on its surface. Since these publications, several other publications in which I participated actively, came out \citep[][]{2014A&A...572A..17M,2016A&A...588A.130M,2017A&A...600L...2C,2017A&A...605A.108M,2017A&A...606L...1W,2018A&A...614A..12M,2018Natur.553..310P}. With increasing observational power and analysis we can now probe the details on the stellar surfaces of different kind of stars. However also the temporal evolution (at different wavelengths) in a key point in the understanding of stellar dynamics: to tackle all the different astrophysical problems related to the evolved stars, today and future interferometers have to challenge the combination of high spectral and spatial resolution as well as the time monitoring on relatively short timescales (weeks/month) of these objects \citep[e.g., ][]{2018A&A...614A..12M}.
\end{itemize}

\textit{\textcolor{blue}{The paper, "Radiative hydrodynamics simulations of red supergiant stars: II. simulations of convection on Betelgeuse match interferometric observations" by Chiavassa et al. 2010, is attached at the end of the chapter.}}.

\subsection{Evolved stars in the Gaia era: probing distances}

Gaia \citep{2016A&A...595A...1G} is an astrometric, photometric, and spectroscopic space-borne mission. It performs a survey of a large part of the Milky Way. The second data release (Gaia DR2) in April 2018 \citep{2018arXiv180409365G} brought high-precision astrometric parameters (i.e. positions, parallaxes, and proper motions) for over 1 billion sources brighter that $G\approx20$. Among all the objects that have been observed, the complicated atmospheric dynamics of AGB and RSG stars is affect the photocentric position and, in turn, their parallaxes. The convection-related variability, in the context of Gaia astrometric measurements, can be considered as `noise' that must be quantified in order to better characterise any resulting error on the parallax determination. However, important information about stellar properties, such as the fundamental stellar parameters, may be hidden behind the Gaia measurement uncertainty.\\
	\cite{2011A&A...528A.120C,2018arXiv180802548C} explored the effect of convection-related surface structures on the photocentre to estimate its impact on the Gaia astrometric measurements. The surface of the deep convection zone has large and small convective cells. The visible fluffy stellar surface is made of shock waves that are produced in the interior and are shaped by the top of the convection zone as they travel outward \citep{2017A&A...600A.137F}. In addition to this, at the wavelengths in Gaia $G$-band (Fig.~\ref{fig:timescales}, top panel, for the RSGs; and bottom panel for the AGBs), molecules (e.g., TiO) produce strong absorption. Both effects modify the position of the photocentre and cause temporal fluctuations during the nominal 5 years of the Gaia mission.
	
\begin{figure}[!h]
\centering
\begin{tabular}{c}
\includegraphics[angle=0,width=0.95\hsize]{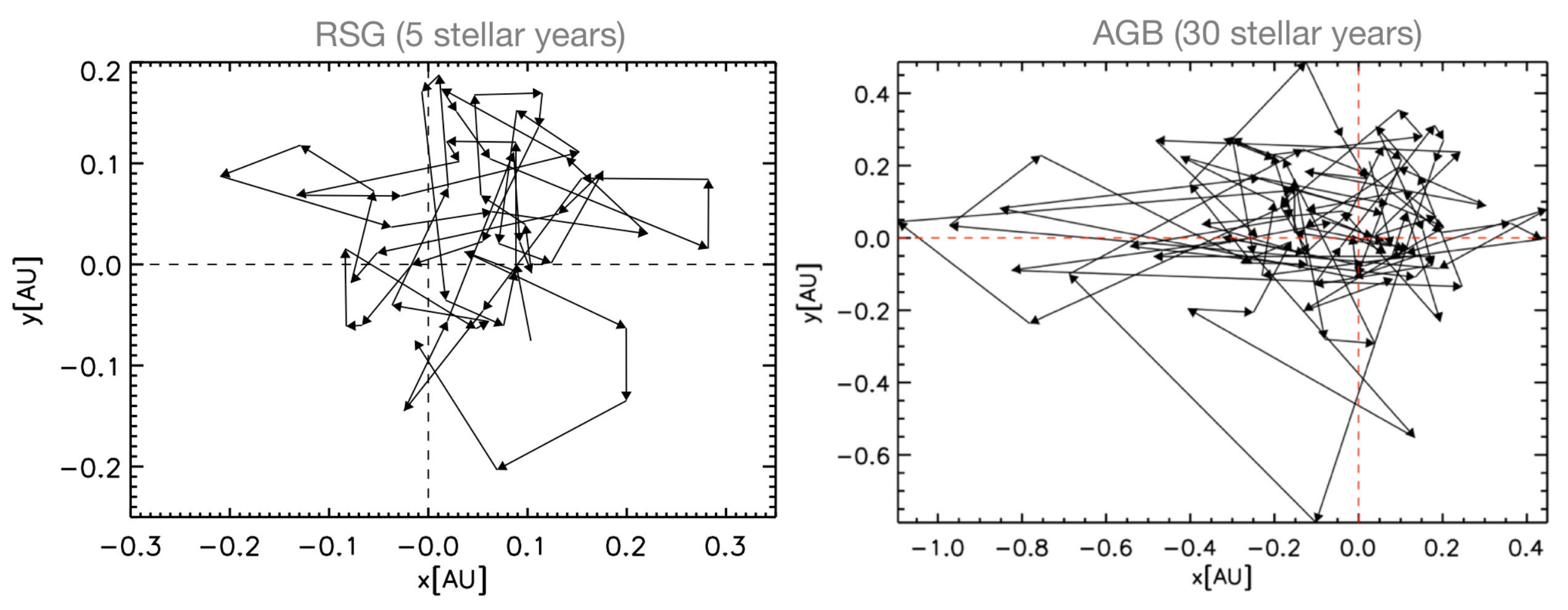}
\end{tabular}
\caption[Photocentre position and Gaia]{Photocentre position computed from 3D RHD simulation of a RSG \citep[\emph{left panel}, the radius is $\sim$4 AU][]{2011A&A...528A.120C} and an AGB \citep[\emph{right panel}, the radius is 1.87 AU][]{2018arXiv180802548C} star. The different snapshots are connected by the line segments; the total time covered is reported above each panel and the snapshots are $\sim$23 days apart. The dashed lines intersect at the position of the geometrical centre of the images.}
\label{fig:agb_rsg_gaia}
\end{figure}

	The position of the photocentre for each snapshot of a simulation (i.e. as a function of time) can be computed as the intensity-weighted mean of the $x-y$ positions of all emitting points tiling the visible stellar surface according to\\
\begin{eqnarray}
P_x=\frac{\sum_{i=1}^{N} \sum_{j=1}^{N} I(i,j)*x(i,j)}{\sum_{i=1}^{N} \sum_{j=1}^{N} I(i,j)} \\
P_y=\frac{\sum_{i=1}^{N} \sum_{j=1}^{N} I(i,j)*y(i,j)}{\sum_{i=1}^{N} \sum_{j=1}^{N} I(i,j)},
\end{eqnarray}
where $I\left(i,j\right)$ is the emerging intensity for the grid
point $(i,j)$ with coordinates $x(i,j)$, $y(i,j)$ of the
simulation, and $N$ is the total number of grid points in the simulated box.  In the presence of surface brightness asymmetries, the photocentre position will not coincide with the barycentre of the star and its position will change as the surface pattern changes with time. This is displayed in the photocentre excursion plots of Fig.~\ref{fig:agb_rsg_gaia}. The fact that $\langle P_x\rangle$ and $\langle P_y\rangle$ do not average to zero means that the photocentre tends not to be centred most of the time on the nominal centre of the star, because of the presence of a large convective cell. 

\marginnote{\footnotesize \it First Gaia result on AGBs: correlation between the photocentre displacement and the pulsation}While \cite{2011A&A...528A.120C} denoted the first predictions for RSG stars, in \citep{2018arXiv180802548C},  we showed that the convection-related variability accounts for a substantial part of the Gaia DR2 parallax error of a sample of semi-regular variables. In addition to this, we presented evidence for a correlation between the mean photocentre displacement and the stellar fundamental parameters: surface gravity and stellar pulsation. This is the first Gaia result on AGB' stars (INSU press release news in November 2018). \\

 \textit{\textcolor{blue}{The paper, "Heading Gaia to measure atmospheric dynamics in AGB stars" by Chiavassa et al. 2018, is attached at the end of the chapter.}}.

\invisiblesection{\it Attached paper: \textbf{main sequence stars}, 3D spectral stellar library and Gaia. \cite{2018A&A...611A..11C}}
\includepdf[pages={1-16}]{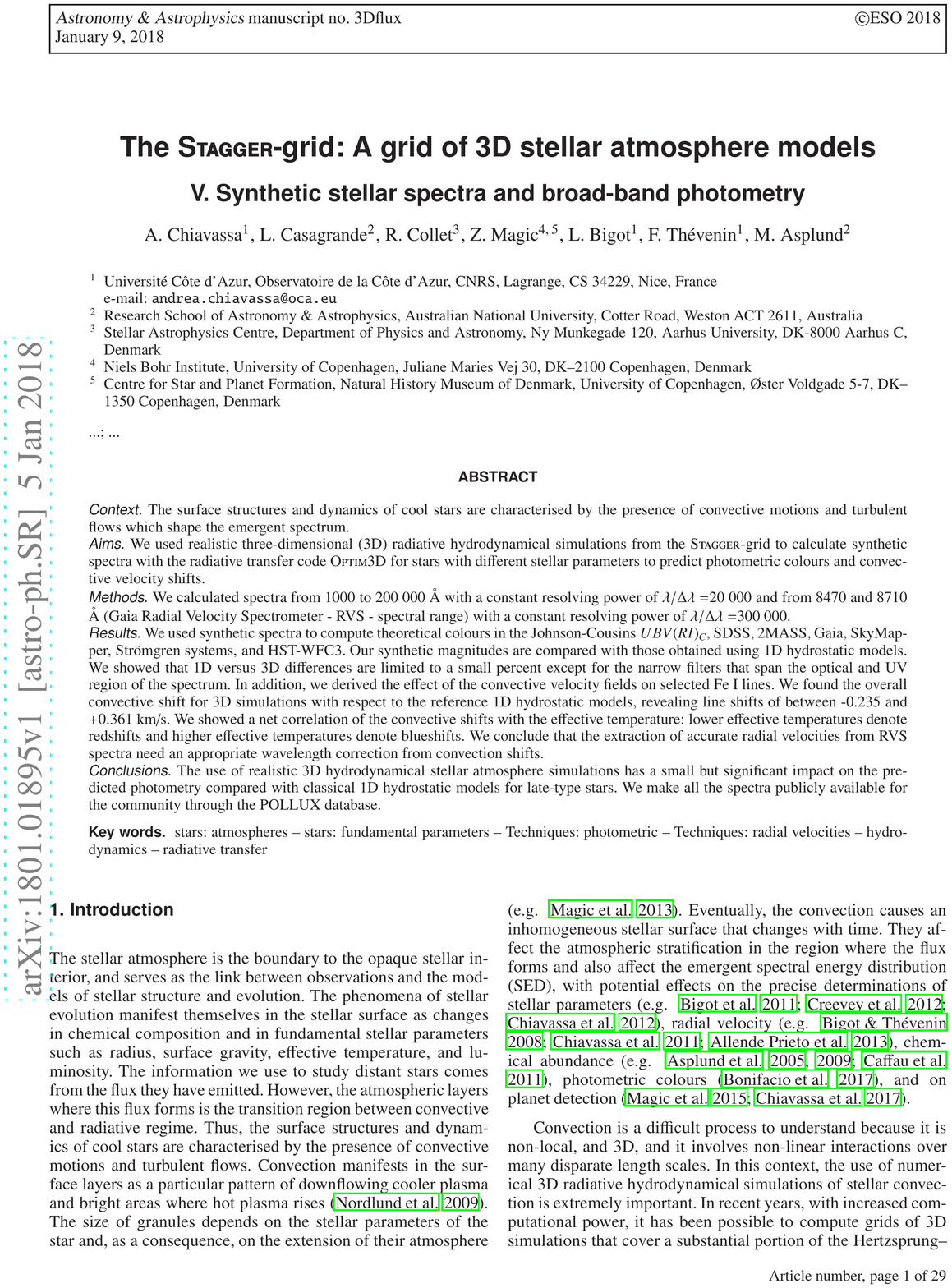}
\invisiblesection{\it Attached paper: \textbf{main sequence stars}, planet synthetic transits with stellar granulation. \cite{2017A&A...597A..94C}}
\includepdf[pages={1-12}]{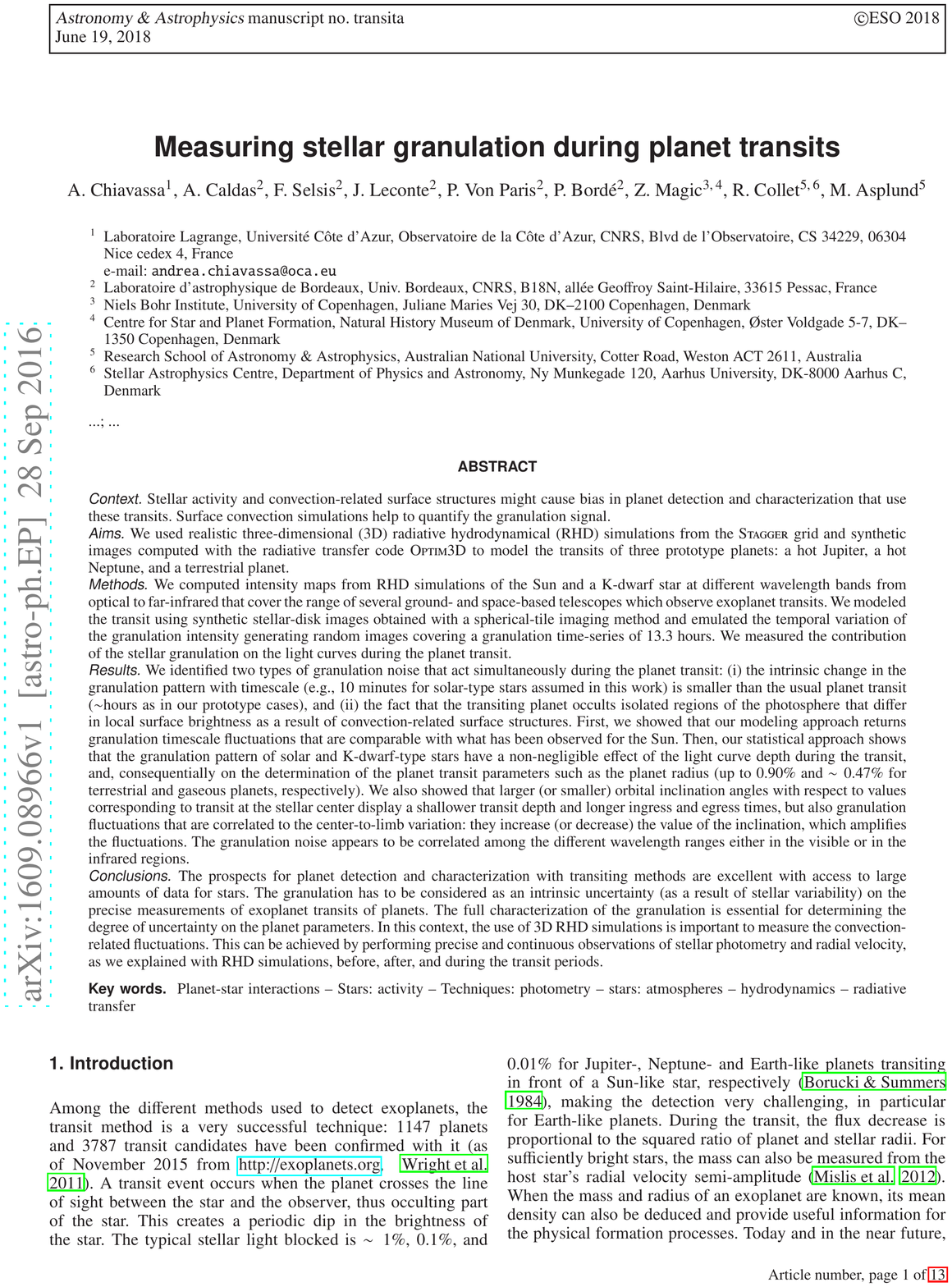}
\invisiblesection{\it Attached paper: \textbf{evolved stars}, characteristic size of convection. \cite{2010A&A...515A..12C}}
\includepdf[pages={1-10}]{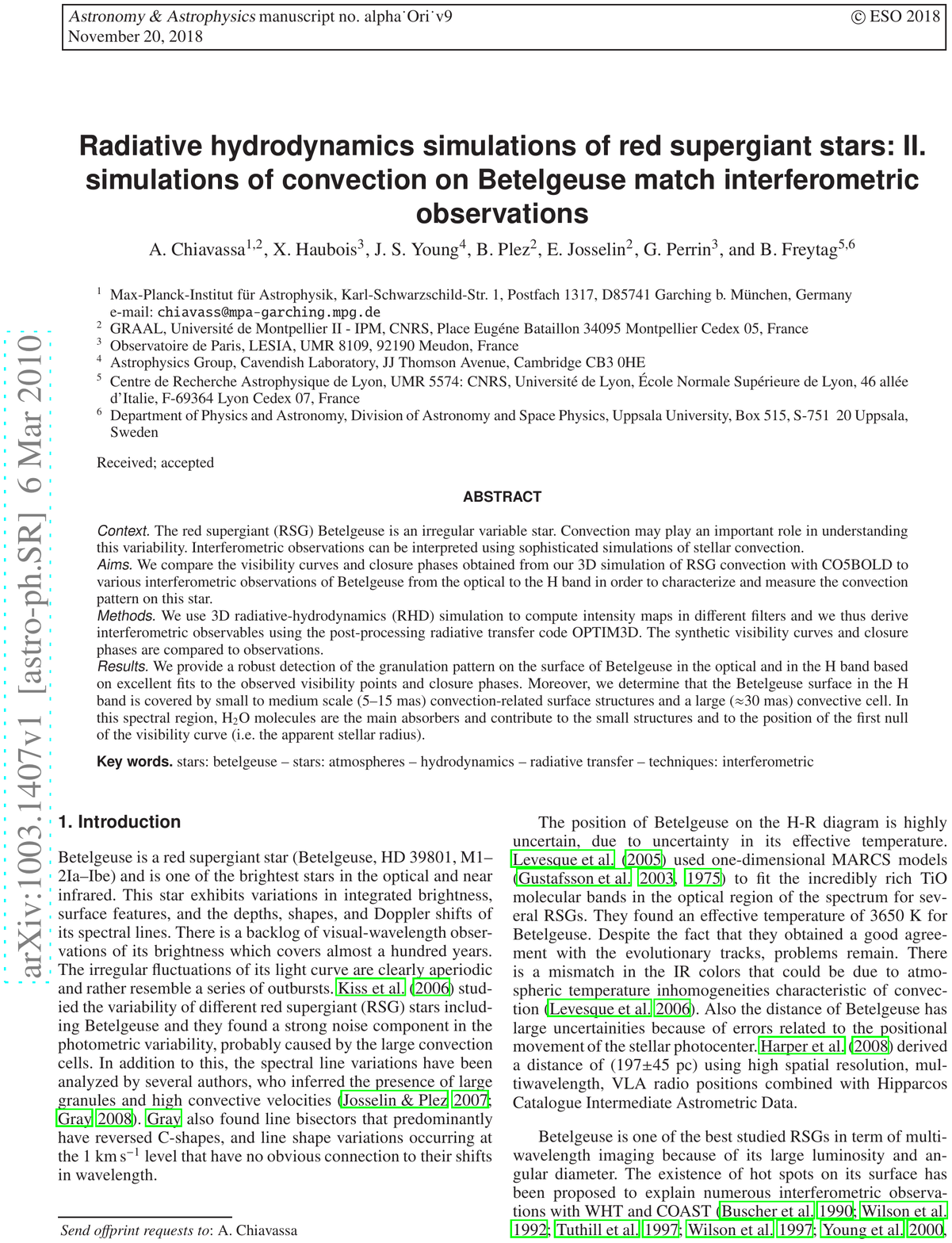}
\invisiblesection{\it Attached paper: \textbf{evolved stars}, characteristic size of convection. \cite{2010A&A...515A..12C}}
\includepdf[pages={1-6}]{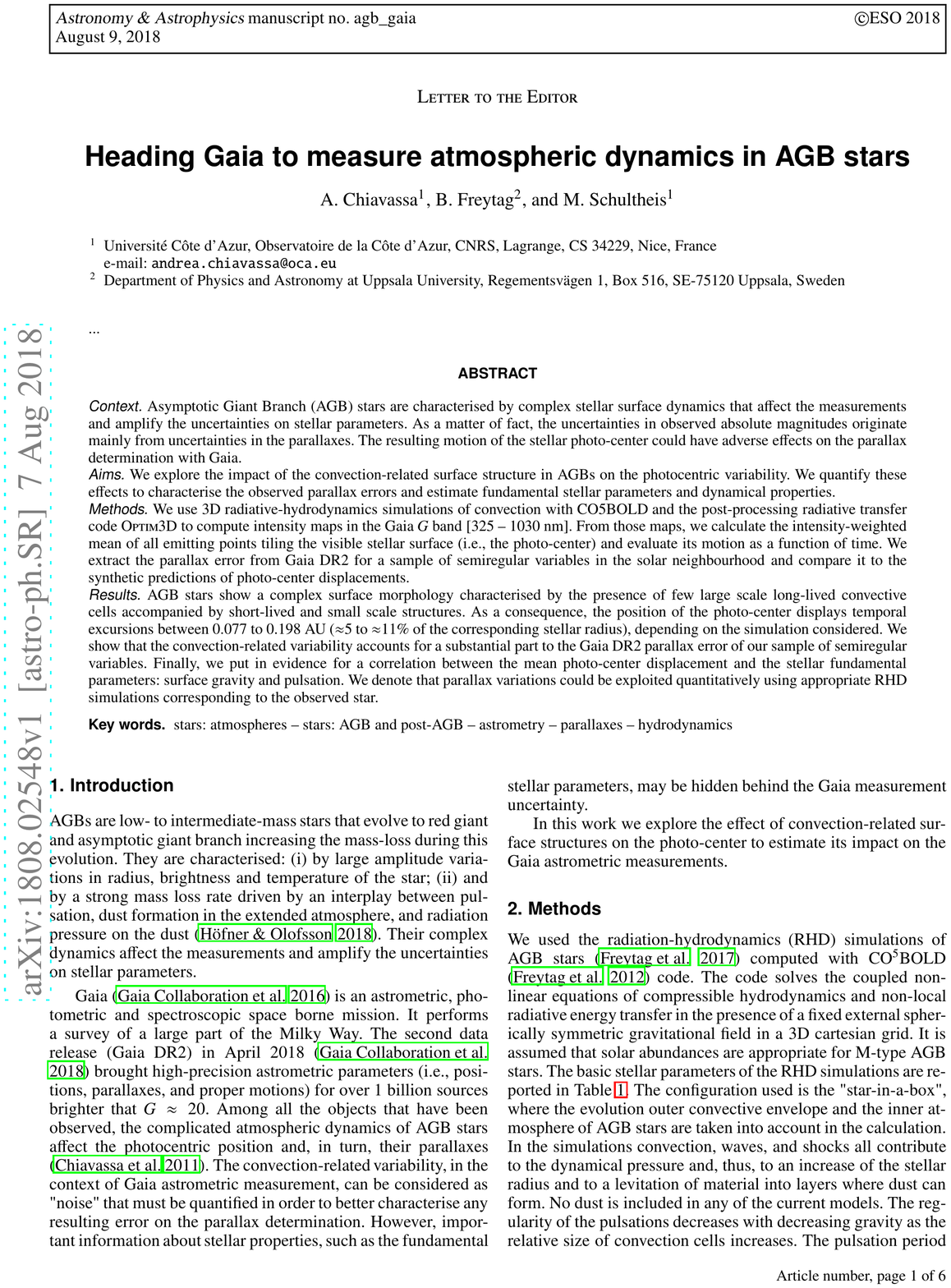}

\clearemptydoublepage

\chapter{Perspectives \label{ch:intro}}

\begin{figure}[!h]
   \centering
  \begin{tabular}{c}
  \includegraphics[width=0.65\hsize]{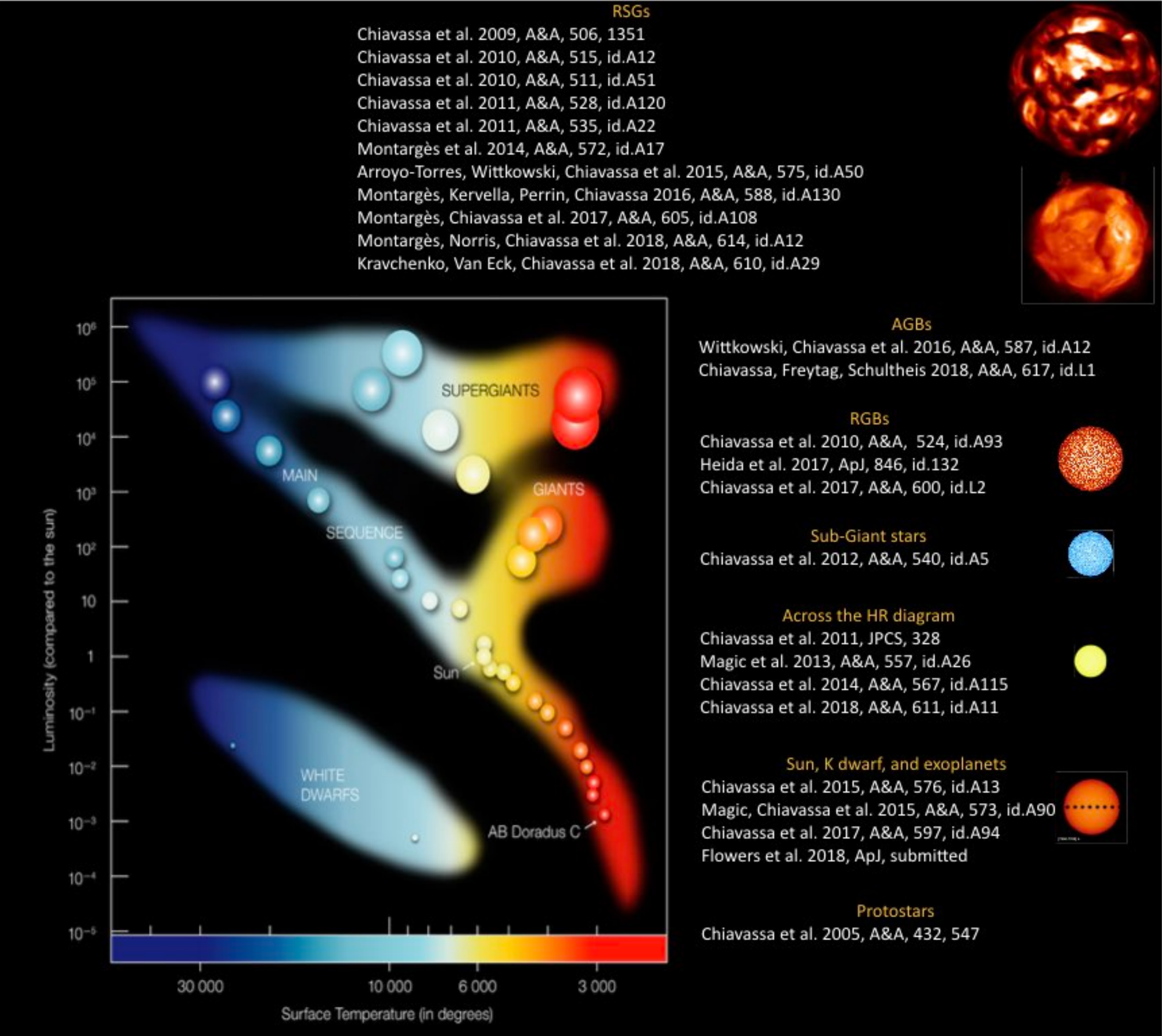}
 \end{tabular}
\caption[Resume of my principal publications]{Resume of my principal publications across the H-R diagram, in which I contributed substantially.}
              \label{fig:maincontrib}%
\end{figure}

Fig.~\ref{fig:maincontrib} displays my principal works for different astrophysical problems I described in previous chapters. On purpose, I indicated only the paper where I largely contributed. In the following sections, I will present where I will drive my research in the coming years.  

\section{HoRSES: High Resolution Spectroscopy for Exoplanets atmospheres and their Stars}

\lettrine[lines=3,nindent=4pt]{T}{he} search for signs of life elsewhere in the Universe requires the remote detection of molecules in the atmospheres of exoplanets. Progress with high-resolution spectroscopy with ground-based instruments has led to detections of atomic and molecular species in the atmospheres of hot giant exoplanets \citep[e.g., ][]{2015A&A...577A..62W,2018A&A...615A..16B}. From the Doppler shift of the planet spectral lines, it has been possible to constrain atmospheric winds \citep{2016ApJ...817..106B,2018arXiv181006099F}, planet rotation \citep{2014Natur.509...63S}, and the orbital inclination of non-transiting planets \citep{2012Natur.486..502B}. Current detections have also the potential to constrain the universal mechanism for planet formation \citep{2016ApJ...833..203P}.

The advent of new high-resolution spectrographs at large and medium-size facilities (CRIRES+, GIARPS, SPIRou, IGRINS, etc...) with unprecedented throughput and spectral range will extend the sample of exoplanets that can be targeted with this technique towards cooler and smaller planets. Given the high degree of complementarity between high-resolution spectroscopy (i.e, ground base instruments) and with low-resolution spectroscopy (space-borne instruments) \citep{2017ApJ...839L...2B}, the synergy between the two techniques will be crucial for the next stage of comparative exo-planetology, especially on the targets found by the TESS mission. When finally implemented at the European Extremely Large Telescope, high-resolution spectroscopy will have the potential to identify biomarkers in the atmospheres of Earth analogues \citep{2014ApJ...781...54R}. \\
However, the planet-hosting stars are covered with a complex and stochastic pattern associated with convective heat transport (i.e., granulation). The convection-related structures have different sizes, depth and temporal variations compared to the stellar type concerned (Fig.~\ref{fig:3Dgran}). The resulting stellar activity, associated to other phenomena such as magnetic field and/or rotation, bias the characterization and detection of exoplanetary signals \citep{2017A&A...597A..94C}. It is of paramount importance to quantify and eliminate convective stellar noise in high spectral resolution observations of exoplanet atmospheres in order to have a solid and unequivocal detection of their chemical composition, planet rotation velocity and winds as well as their thermal structure. This can be applied now in giant planets, but, in the TESS era, it will be used for objects of comparable size to the Earth. 

\begin{figure}[!h]
   \centering
  \begin{tabular}{c}
  \includegraphics[width=0.9\hsize]{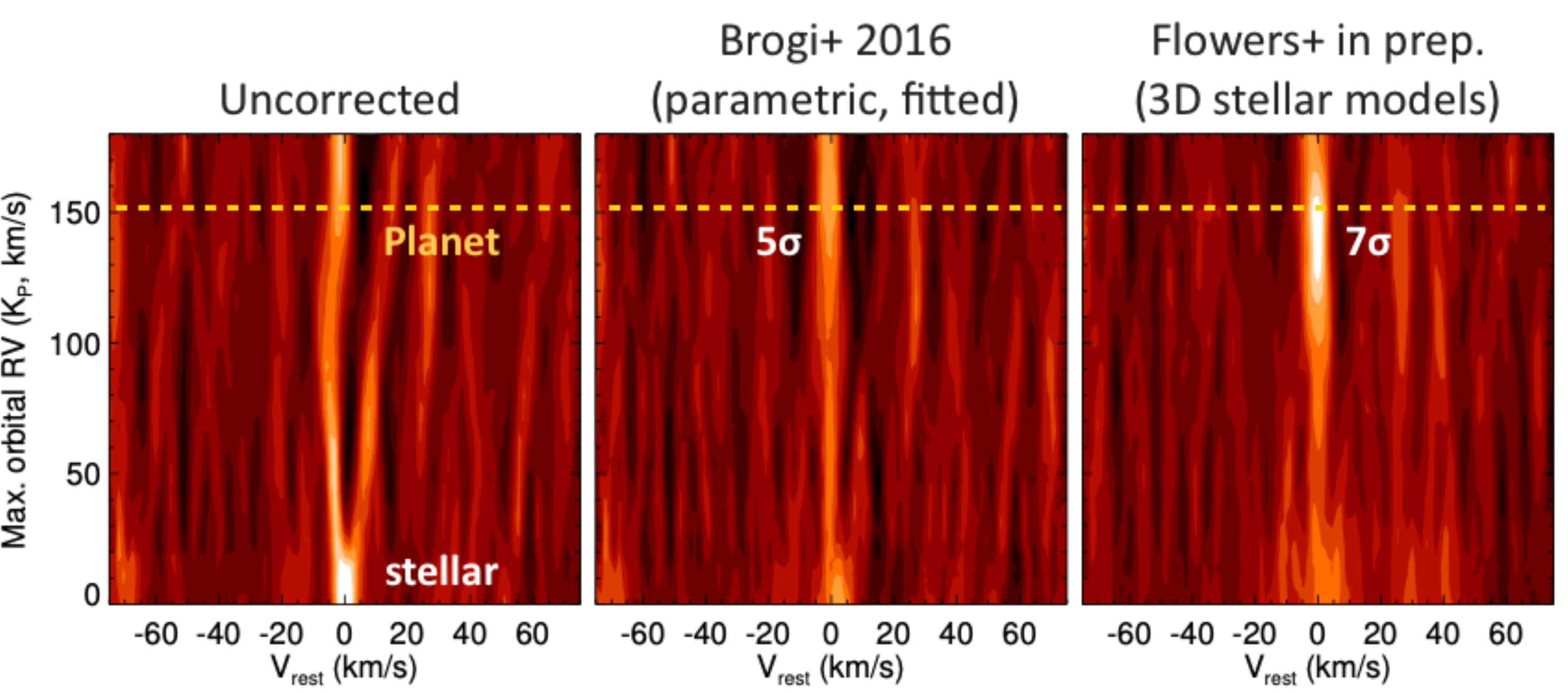}
 \end{tabular}
\caption[Exoplanet atmospheres at high spectral resolution]{Two-dimensional significance maps of the explored planet parameter orbital radial-velocity semi-amplitude ($K_P$). While in left panel the expected planet value (yellow dashed line) is completely hidden by stellar signal, in central panel (paramentric approach) and right panel (3D RHD simulation correction) the planet orbital radial-velocity is recovered with an higher significatively when a proper stellar correction is applied.}
              \label{fig:ccf}%
\end{figure}

In this context, I started in 2017 a collaboration with M. Brogi (University of Warwick, UK), one of the worldwide know expert in high spectral resolution for exoplanetary atmospheres. Together, we organized in October 2018 the first conference on this topic\footnote{https://horse.sciencesconf.org and \#NICEEXOPLANETS}. The event was a real (somehow unexpected) success where we managed to gather worldwide experts in the topic of exoplanet atmospheres and stellar physics meet together around the high spectral resolution spectroscopy. A follow up meeting is expected to take place in 2 years from now. \\
	 Our collaborations is concretising very recently with a first publication \citep[][ submitted to ApJ]{2018arXiv181006099F} where, for the first time, we used 3D RHD simulation to match CO lines of the well known hot Jupiter HD 189733b observed in transmission spectroscopy with CRIRES. A second paper (Brogi $\&$ Chiavassa) is preparation for a second non-transiting exoplanet). In planetology community, using a full consistent model for the stellar component (i.e., including granulation effect on spectral line shape, depth, and velocity) is a major improvement from previous studies, which instead relied on parameterizing an average stellar line profile through micro- and macro-turbulence\citep{2012Natur.486..502B}, and was inadequate to reproduce the complicated velocity fields in the convective envelope of the star (Fig.~\ref{fig:ccf}).
	 
	 The doorway is now open and several studies are already on going or about to start. They concern the deeper analysis of several detections of atomic lines in hot Jupiters (Sodium, Potassium, Helium) as well as Molecules and their relative abundances (CO, H$_2$O, CH$_4$,...). \\
	 The ability to obtain reliable measurements of planetary properties (i.e., at high spectral resolution), including detailed characterisation of the star-host, will be the cornerstone on which the characterisation of habitable planets will be based.

\clearemptydoublepage

\section{Improving 3D simulations of RGSs towards the solution of the mass-loss problem}

\begin{figure}[!h]
   \centering
  \begin{tabular}{c}
  \includegraphics[width=0.9\hsize]{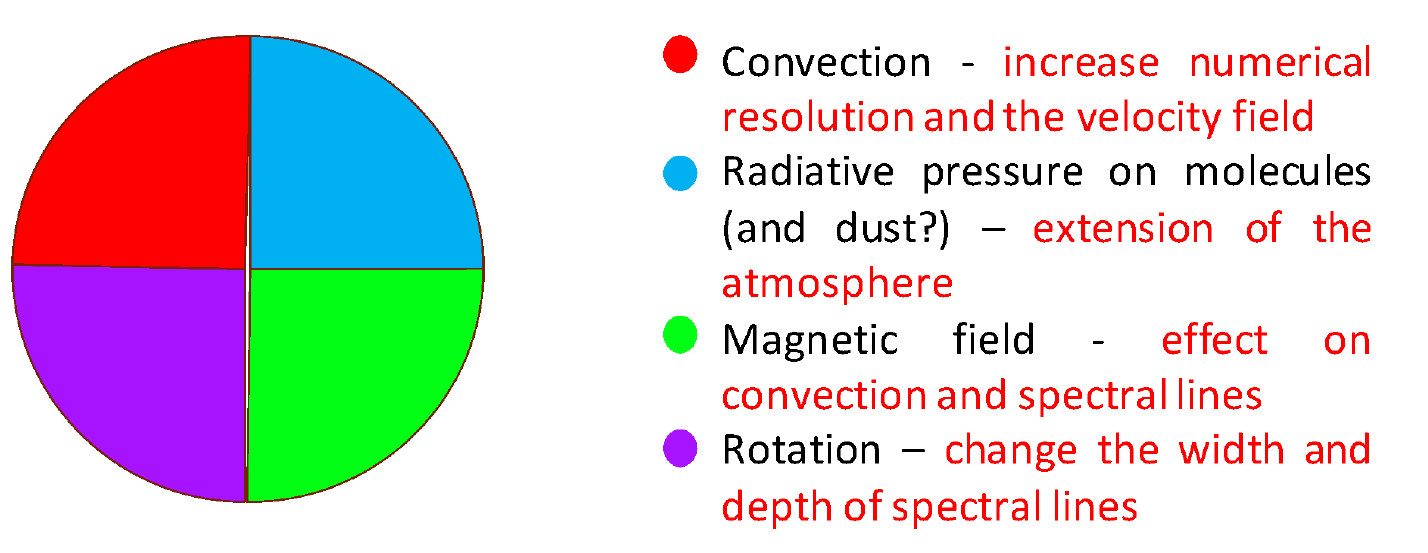}
 \end{tabular}
\caption[The four main limitations of 3D RHD simulations of RSG stars]{The four main limitations of 3D RHD simulations of RSG stars to be solved in order to provide a quantitative response to the problem of mass loss that affects the chemical evolution of galaxies. In red the  effect on observables and/or the simulations}
              \label{fig:limitations}%
\end{figure}

\lettrine[lines=3,nindent=4pt]{I}{presented} the simulations of the red supergiants as well as the code CO$^5$BOLD. Despite the very satisfactory comparisons with observations, recent advances in instrumental techniques in interferometry (PIONIER and AMBER@VLTI), imaging (SPHERE@VLT) and spectroscopy have achieved an astonishing level of accuracy. A number of recent studies, in which I took part, have highlighted the current limitations of 3D simulations that need to be solved in order to provide a quantitative response to the problem of mass loss in red supergiants (Fig.~\ref{fig:limitations}). These four points, in which I will concentrate my research, are the cornerstone for future developments in the field, in particular the magnetic field and radiative pressure.\\ 

\begin{figure}[h!]
\centering
\begin{tabular}{cc}
\includegraphics[angle=0,width=0.6\hsize]{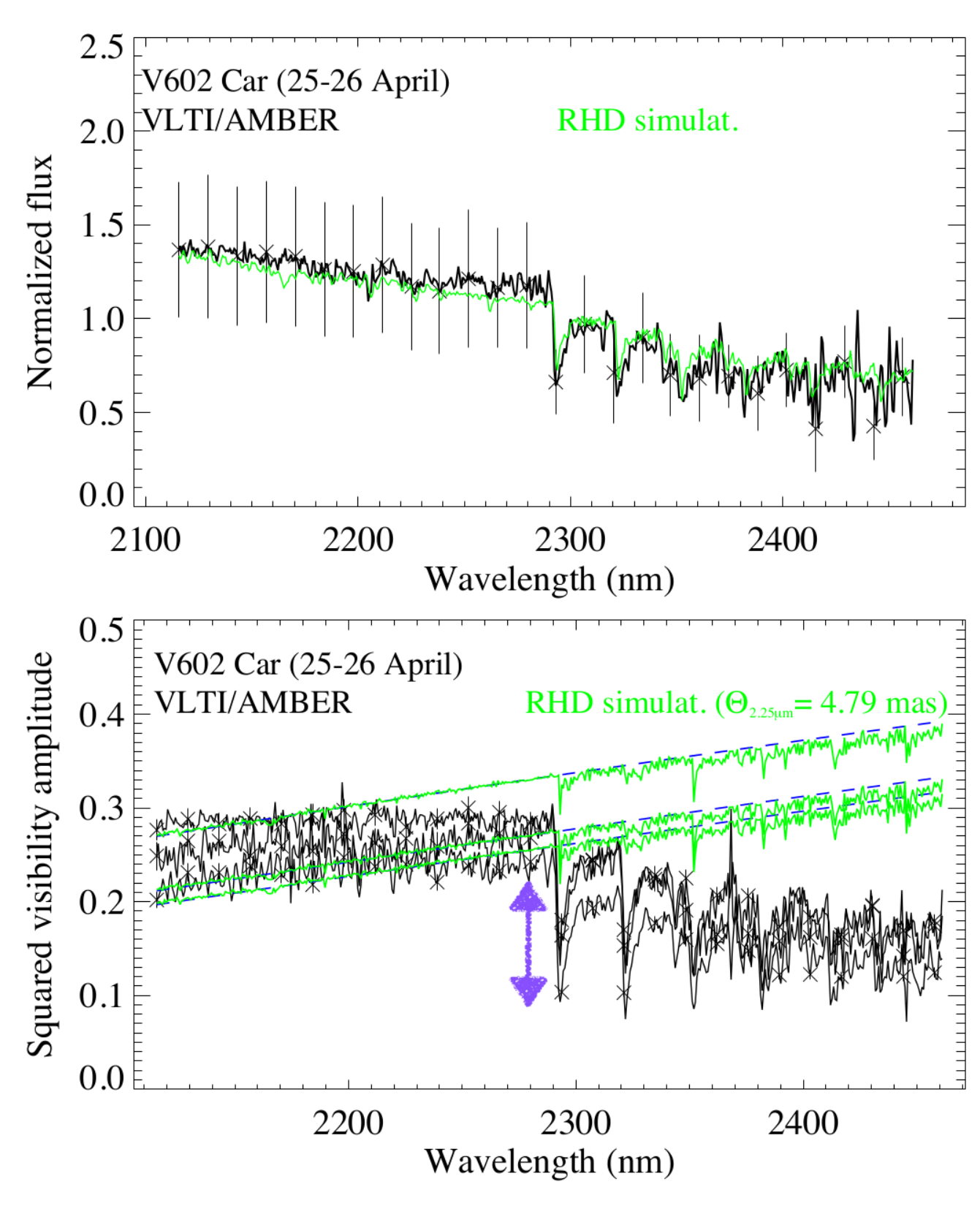} 
\includegraphics[angle=0,width=0.4\hsize]{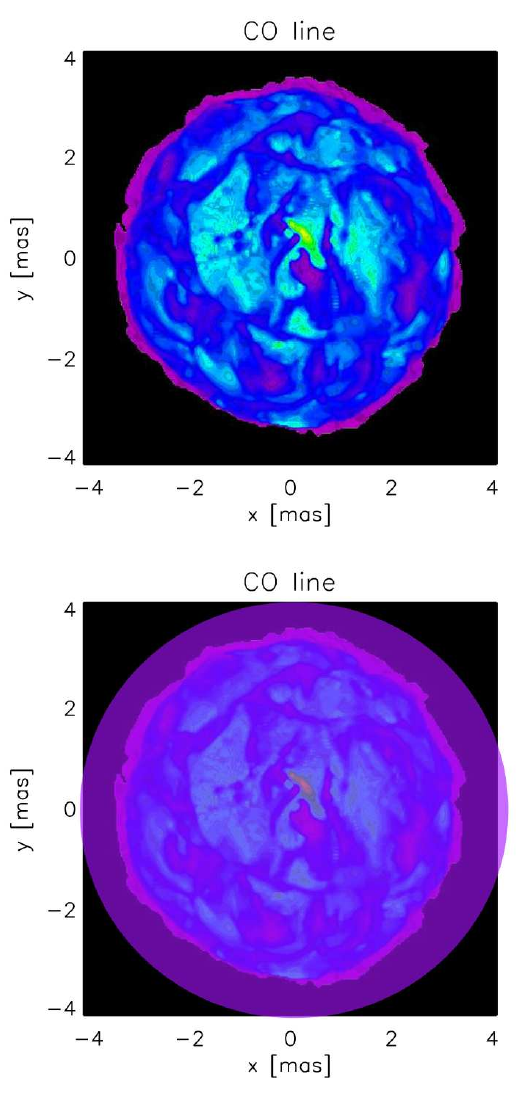} 
\end{tabular}
\caption[Atmospheric extension problem]{\emph{Left column:} Interferometric observations of red supergiants with AMBER (black) compared to 3D simulation predictions (green). While the flux adjustment (top panel) is good, the visibilities show a significant disagreement (bottom panel, purple arrow). \emph{Right column:} Synthetic image in the CO lines at about 2.3 $\mu$m. The image at the top shows the result of the 3D atmosphere calculation, while the purple shadow in the image at the bottom displays the expected atmospheric extension needed to explain the difference in visibility \citep{2015A&A...575A..50A}}
\label{rad_press}
\end{figure}

\begin{enumerate}
\item \emph{The numerical resolution}. It affects the source function due to the lack of spatial resolution at about the optical depth of $\tau=1$ (i.e., where the flux is formed). As a result, the emerging intensity may show, in some cases, extreme peaks of brightness with respect to the to adjacent areas. Attempts have been made to solve this problem by interpolating the source function in 
CO$^5$BOLD, but they have caused numerical instabilities, and in {{\sc Optim3D}}. In addition to this, numerical resolution determines also the resolving power of the structures on the stellar surface. This leads to an under estimation of the simulation's velocities that affects the gas levitation and the width of spectral lines (K. Kravchenko's thesis). The current solution is to increase the number of
points at $1000^3$ or more, which implies a more intensive and time consuming use of cluster computers. 
\item \emph{The radiative pressure}. Recent comparisons of simulations to interferometric observations with AMBER have shown that the extension of the observed red supergiant atmospheres is not explicable by current models \citep[Fig.~\ref{rad_press}, ][]{2015A&A...575A..50A}. This vision was confirmed by the observations taken with SPHERE in the optical where the extension of 3D atmospheres is too small \citep{2016A&A...585A..28K}. In addition to this, the images also show a dust shell in the 3 stellar rays that could be the consequence of a photospheric ejection. The inclusion of radiative pressure in the simulations should help the gas to levitate in the outermost layers of the atmosphere, where opacity is not negligible (e.g., TiO molecules) and explain (at least in part) the mechanism of mass loss: radiative pressure on molecules.
\item \emph{The magnetic field}. The presence of a magnetic field in stars is intimately linked to the convection across the stellar photosphere. A typically magnetic field results in the increase of atmospheric velocities and higher temperatures in the chromosphere. As a consequence, the overall structure of the stellar atmosphere is affected. In the case of evolved stars, local dynamos are expected to appear in correspondance of the large convective cells \citep{2002AN....323..213F}. Longterm spectropolarimetric observations \citep[e.g., those obtained with NARVAL at the Pic du Midi, ][]{2010A&A...516L...2A, 2018A&A...615A.116M} show the detection of a signal in the spectral lines and measure an average longitudinal magnetic field of a few Gauss. The introduction of the magnetic field into 3D RHD simulations is under development and will be of paramount importance for the interpretation of observations.
\item \emph{The stellar rotation.} Observations indicate that synthetic spectral lines are narrower and deeper than those observed \citep[e.g., ][]{2013EAS....60...85L}. This may be caused by the low numerical resolution of point (1), but it may be also caused by the lack of a rotational velocity. \cite{1998AJ....116.2501U} found a angular rotation velocity between 2.0 and 2.5 km/s (i.e., a projected equatorial velocity of 5.0 km/s) that could (partially) contribute to the loss of mass. Recently, this result was confirmed by \cite{2018A&A...609A..67K}, with ALMA observations, who found a projected equatorial velocity of $\sim5.47$ km/s. Recently, I computed a series of 3D RHD simulations with different projected equatorial velocities ranging from 2.0 to 6.0 km/s, which corresponds to a rotation period of 60 to 20 years, respectively. A record for this type of simulation that required an intensive use of computing resources. These simulations will be exploded soon. 
\end{enumerate}

%
%
%

\clearemptydoublepage
\chapter*{Abbreviations}

\begin{table}[!h]
\footnotesize
\begin{tabular}{ll} 
1D & One-dimensional models\\
3D & Three-dimensional models\\
AGB &  Asymptotic Giant Branch\\
 box-in-a-star & local RHD simulations \\
CLV & Center to limb variations \\
CO$^5$BOLD & COnservative COde for the COmputation of COmpressible COnvection in a BOx of L Dimensions, 1=2,3" \\
 ESO		&	European South Observatory\\
 LTE 		& Local Thermodynamic Equilibrium\\
 H-R           & Herzsprung-Russel\\
 HST-WFC3  & photometric system \\
 IR		&	InfraRed\\
 ISM           &      InterStellar Medium\\ 
 UBV(RI)$_{\rm{C}}$  &   Johnson-Cousins, photometric system \\
 MARCS & 1D, hydrostatic, plane-parallel and spherical LTE model atmospheres \\
 2MASS  & photometric system \\
 {{\sc Optim3D}}  &  Multidimensional pure LTE radiative transfer code \\
 RHD         &       Radiative HydroDynamical\\
 RGB         &       Red Giant Branch stars \\
 RSG         &       Red Supergiant stars\\
 SED         &  	Spectral Energy Distribution\\
 SDSS      &   Sloan Digital Sky Survey, photometric system \\
 SkyMapper & Photometric systems \\
 Stagger code & multipurpose, radiative-magnetohydrodynamics code \\
 star-in-a-box & global RHD simulations \\
 Str\"omgren  & Photometric system \\
 TURBOSPECTRUM & Code for 1D spectral synthesis \\
UV-plane & The Fourier plane with u,v coordinates related to x, y ones in the sky \\
\\
 UV 		&	UltraViolet\\
 VALD       &      Vienna Atomic Line Database\\
\end{tabular}
\end{table}

\clearemptydoublepage

\bibliographystyle{aa}
\bibliography{biblio.bib}
 


\clearemptydoublepage

\end{document}